\pdfoutput=1

\documentclass[12pt]{article}
\usepackage[
top = 2.5cm, 
bottom = 2.5cm,  
left = 2.55cm,  
right = 2.55cm]{geometry}

\usepackage{latexsym} 
\usepackage{verbatim}
\usepackage{tikz}
\usetikzlibrary{matrix}

\usepackage{color}
\usepackage{graphicx,amssymb,amsfonts,amsmath,amssymb,amscd,amstext, mathrsfs}
\usepackage{graphicx} 
\usepackage{bbm}
\usepackage{dsfont}
\usepackage{xcolor}
\usepackage[colorlinks=true, citecolor=black, linkcolor=black, allcolors=black]{hyperref}   
\usepackage{amsmath}
\usepackage{empheq}
\usepackage{hyperref}

\newlength\dlf

\def\be{\begin{eqnarray}}
\def\ee{\end{eqnarray}}
\def \bea {\begin{equation}}
\def \eea {\end{equation}}

\newcommand{\CO}{{\cal O}}
\newcommand{\CV}{{\cal V}}
\def \nn {\nonumber}

\def \ra {\rangle}
\def \rr {\raise.35ex\hbox{\small $\prime$}\kern-.17em{\mbox{\large $\imath$}}}
\def \del {\partial}
\def \dels {\partial\kern-.5em / \kern.5em}
\def \As {{A\kern-.5em / \kern.5em}}
\def \Ds {D\kern-.7em / \kern.5em}

\def \a {\alpha}

\def \b {\beta}

\def \m {\mu}
\def \n {\nu}

\def \th {\theta}

\def\frac#1#2{{#1\over #2}}

   \def \b {\beta}

\newcommand{\<}{\langle}
\renewcommand{\>}{\rangle}

\def \iffa {\iffalse} 

\def \ed  {\end{document}}

  \def \a {\alpha}
  \def \b {\beta}
\def\nn{\nonumber}

\def \ed  {\end{document}}

\def \Hat {\hat}
\def \y {\Hat \rho}
\def \b {\Hat b}
\def \th {\hat{\theta}}

\usepackage{cite}
\begin{document}
\thispagestyle{empty}
\begin{flushright} 

\end{flushright}
\begin{center} 

{\LARGE{\textsc{\bf{Universal Lowest-Twist in CFTs  
\\
\vspace{0.4cm} 
from Holography}}}}
\vspace{1.2cm}  

A. Liam Fitzpatrick and Kuo-Wei Huang
\\ 
\vspace{0.8cm} 
{\it
Department of Physics, Boston University,\\
Commonwealth Avenue, Boston, MA 02215, USA}\\
\end{center}
\vspace{2.2cm}

\noindent{We probe the conformal block structure of a scalar four-point function in $d\geq2$  conformal field theories   
by including higher-order derivative terms in a bulk gravitational action. 
We consider a heavy-light four-point function as the boundary correlator at large central charge.
Such a four-point function can be computed, on the gravity side, as a two-point function of the light operator 
in a black hole geometry created by the heavy operator. 
We consider analytically solving the corresponding scalar field equation in a near-boundary expansion 
and find that the multi-stress tensor conformal blocks are insensitive to the horizon boundary condition. 
The main result of this paper is that the lowest-twist operator product expansion (OPE)    
coefficients of the multi-stress tensor conformal blocks are universal:  
they are fixed by the dimension of the light operators and the ratio between 
the dimension of the heavy operator and the central charge $C_T$. 
Neither supersymmetry nor unitary is assumed.  Higher-twist coefficients, on the other hand, generally are not protected.
A recursion relation allows us to efficiently compute universal lowest-twist coefficients.  
The universality result hints  at 
the potential existence of a higher-dimensional Virasoro-like symmetry near the lightcone.    
While we largely focus on the planar black hole limit in this paper, we include 
some preliminary analysis of the spherical black hole case in an appendix.  
}

\newpage
\tableofcontents

\newpage

\addtolength{\parskip}{0.5 ex}
\jot=1.5 ex

\setcounter{equation}{0} 
\setcounter{footnote}{0}
\setcounter{section}{0}
\renewcommand{\theequation}{1.\arabic{equation}}
\section{Introduction and Summary}

The AdS/CFT correspondence \cite{Maldacena:1997re, Witten:1998qj, Gubser:1998bc} provides powerful insights  from gravity in anti-de Sitter (AdS) to conformal field theory (CFT) and vice versa.  A remarkable amount of the usefulness of the correspondence does not depend on detailed knowledge of any specific pair of dual theories, but rather follows from the fact that many of the important properties of one side of the correspondence are automatically built into the other side as well, but in a different framework.  An important example is that gravity -- the existence of a massless, spin-2 particle -- is built into the CFT through the stress tensor and its Ward identities.  Another is that crossing symmetry of CFT correlators, a highly nontrivial constraint, is built into Witten diagrams in the bulk theory \cite{Heemskerk:2009pn}.  
In such cases, it is somewhat arbitrary to say whether one is using the CFT or gravity side   
since the duality is merely acting as two different languages for the same physics, closely analogous to the relation between S-matrices and their Lagrangians  in flat space.  

In two-dimensions, where the CFT formulation has been especially useful, the existence of the Virasoro symmetry might be
difficult to discover on the gravity side \cite{Brown:1986nw}, but once discovered it leads to a computationally powerful algebraic description of many  gravitational effects.  
In particular, when applied to four-point functions of local CFT operators, the irreducible representations of the algebra, known as Virasoro Conformal Blocks, capture the thermal properties of black holes, the information paradox associated with late-time decay of correlators in semiclassical gravity \cite{Maldacena:2001kr,Fitzpatrick:2014vua,Fitzpatrick:2015zha}, the universal properties of Renyi and entanglement entropy 
\cite{Hartman:2013mia,Asplund:2014coa, Chen:2016kyz, Chen:2016dfb} leading to a proof of Ryu-Takayanagi formula \cite{Ryu:2006bv} in  AdS$_3$/CFT$_2$,  the maximal growth of chaos in gravity \cite{Roberts:2014ifa,Anous:2019yku}, the nonperturbative resolution of some perturbative violations of unitarity \cite{Fitzpatrick:2016ive,Chen:2017yze}, and more, all without appealing to a gravitational Lagrangian. 
See also \cite{Banerjee:2016qca, Hijano:2015qja, Hijano:2015rla, Hartman:2014oaa, Anous:2016kss, Kulaxizi:2018dxo,Fitzpatrick:2015foa,Anand:2017dav,Chen:2017dnl,Fitzpatrick:2015dlt} for related discussions.

One of the main motivations of the analyses in this paper is to try and generalize the Virasoro vacuum blocks of $d=2$ CFTs to higher dimensions.  A natural  higher dimensional analogue of a two-dimensional Virasoro vacuum block is the contribution to a four-point function $\< \CO_1 \CO_1 \CO_2 \CO_2\>$  in the $\CO_1 \CO_1 \rightarrow \CO_2 \CO_2$ channel from  
all  operators made from products of stress tensors, which we will refer to as ``$T^n$s''.
An immediate issue one has to deal with is that in $d>2$, the  $T^n$s are no longer controlled by the conformal algebra. 
In fact, due to the operator mixing, it is 
ambiguous what operators one should identify as the products of stress tensors. 
One may take an infinite central charge $C_T$ limit where the stress tensor and its products form a Generalized Free Field (GFF) theory subsector of the full theory, and 
 $T^n$s have a canonical definition.\footnote{Following  \cite{Osborn:1993cr,Dolan:2000ut}, $C_T$ is the coefficient of the $T^{\mu\nu}$ two-point function.
In $d=4$, $C_T= {40\over \pi^4} c$ in terms of $c$ in the trace anomaly $ \langle T^\mu_\mu \ra= {c\over 16 \pi^2} W^2_{\mu\nu\lambda\rho}+...$ where $W_{\mu\nu\lambda\rho}$ is the Weyl tensor. In $d=2$, $C_T=2c$.
}  
However, the contributions from $T^n$s (in a conformal block decomposition) to boundary correlators 
in the theory still are sensitively
theory-dependent, involving more and more parameters as  higher and higher powers of the stress tensor are included.  So it is not clear what, if anything, one could compute about such multi-stress-tensor contributions -- dual to multi-graviton effects -- in a model-independent way.

From this point of view, effective Lagrangians for gravity in AdS are a useful lab for investigating potentially universal features of CFT.  By dialing the parameters of the bulk Lagrangian, one can sweep out large classes of  CFT  data, i.e. the scaling dimensions  and the Operator Product Expansion (OPE) coefficients of local boundary operators, consistent  
with standard CFT axioms within some regime of validity \cite{Heemskerk:2009pn, Fitzpatrick:2010zm,Fitzpatrick:2011hu,Unitarity,AdSfromCFT}. 
Anything that {\it is} universal in CFT will be invariant under changes of the bulk Lagrangian, and moreover 
may be computed perturbatively on the gravity side.  
At large central charge $C_T$, a particularly interesting set of boundary correlators are the ``heavy-light'' four-point functions, of two  light  operators $\CO_L$ with dimensions $\Delta_L$ much less than $C_T$ and two heavy  operators $\CO_H$ with dimensions $\Delta_H$ parametrically the same as $C_T$, to compensate for the large $C_T$ suppression of the stress tensors.  
In $d=2$, the Virasoro conformal block for heavy-light four-point functions at large  $C_T$ was originally computed using CFT techniques \cite{Fitzpatrick:2014vua,Fitzpatrick:2015zha}, but it could just as easily have been read off from a semiclassical gravity computation.  On the gravity side, the heavy-light four-point function is just a two-point function of the light operator in a black hole background created by the heavy operator, 
 \be
\<\CO_L(x_1) \CO_L(x_2)\>_{\rm BH}  \ ,
 \label{eq:2point}
 \ee
and thus all one has to do is to compute this two-point function 
with the BTZ metric \cite{PhysRevLett.69.1849}  
and extract the contributions from the boundary stress tensors.\footnote{We will describe this derivation more explicitly in subsection \ref{sec:BTZ}. Strictly speaking, because a black hole in AdS is a canonical ensemble, (\ref{eq:2point}) actually thermally 
averages over the heavy operators in the four-point functions. At infinite $C_T$, the thermal average localizes on heavy states of a definite energy, and the $d=2$ conformal algebra implies that the vacuum Virasoro block is the same for different heavy states with the same dimension, so we can still extract the Virasoro vacuum block at infinite $C_T$ from (\ref{eq:2point}) in the way described.  
Moreover, in the infinite temperature limit, which will be of particular 
interest, the canonical ensemble and microcanonical ensemble should agree. 
}  The $d=2$  
CFT description guarantees that the result is independent of the particular gravity theory used to do the computation, but the computation itself is done completely on the gravity side.   

In this paper, we will use a similar strategy in higher dimensions.  That is, we will consider a large class of gravity theories and look for universal contributions to heavy-light correlators within that class.  We will also use the gravitational description to explicitly 
compute these contributions.  The large class we consider is AdS theories that can be written as a scalar field $\phi$ coupled to gravity with {\it arbitrary} 
higher-curvature terms in the action, in the limit of infinite $C_T$.\footnote{Even if there are additional massive fields in the bulk, if their masses are controlled by a free parameter that can be dialed to make them heavy then we can consider a series expansion in inverse powers of their masses.  To all orders in such an expansion, the contributions from the massive fields can be absorbed into the higher-curvature terms in the gravitational action. Thus, our class of theories is larger than it may seem. 
}  
We will focus on the contribution to heavy-light correlators from $T^n$ operators, which are the natural generalization of the $d=2$ Virasoro conformal block.  
Most contributions from $T^n$s will indeed be theory-dependent. However, our main result is that there is a special class of multi-stress tensor operators that is universal.  We will call this class of operators the ``lowest-twist'' $T^n$s.  For any number $n$ of stress tensors inside the product, we define the lowest-twist operators as those with the smallest possible twist for that number of stress tensors.\footnote{Recall that the twist of an operator is just its dimension minus its spin, $\tau = \Delta-J$.} Since each stress tensor raises the dimension by $d$ and raises the spin by at most $2$, the lowest possible twist at each $n$ is 
\be
\tau_{\rm min}(n) = n(d-2) \ .
\ee
 These lowest-twist $T^n$ operators are essentially operators made from stress tensors without contracting any Lorentz indices.    

By ``universal'', we mean that the OPE coefficients, $c_{\rm OPE}$, of the lowest-twist $T^n$s are completely fixed in terms of the following physical CFT data: the dimension $\Delta_L$ of the light operators, the dimension $\Delta_H$ of the heavy operator, and the central charge $C_T$.  In the infinite $C_T$ limit that we consider, the latter two appear only in the combination 
\be
f_0 = \frac{4 \Gamma(d+2)}{(d-1)^2 \Gamma^2(\frac{d}{2})} \frac{\Delta_H}{C_T} \ ,
\label{eq:f0def1}
\ee
in units of the AdS radius of curvature. On the gravity side, the factor $f_0$ is simply defined as the coefficient of 
the first correction to the bulk metric in an expansion near the boundary of AdS; see (\ref{asyfr}) for the precise definition.  
Our universality result can be summarized as
\be
\label{f0sum}
c_{\rm{OPE}}\big(\tau_{\rm min}(n),J\big)= f_0^n {\cal F}_{n,J}(\Delta_L) \ ,
\ee 
where the function ${\cal F}_{n,J}$ is independent of higher-curvature parameters in the bulk action.\footnote{We will derive this statement in the limit of planar black holes, and simply provide some evidence for spherical black holes. In the planar limit, there is only one $T^n$ that contributes at each $n$ and so the $J$ label is somewhat superfluous.}  
Neither supersymmetry nor unitary is assumed.\footnote{For a unitary CFT, $C_T>0$.} The higher-twist OPE coefficients, on the other hand,  can be contaminated by other generally model- and coupling-dependent parameters in a CFT.

Operators with low twist are interesting for a number of reasons, mainly because operators with lower twist produce larger contributions in the limit that some of the operators in a correlator approach each other's lightcone. 
For instance, in $d=4$, 
the leading behavior of a conformal block for a four-point function $\< \CO_L \CO_L \CO_H \CO_H\>$ near the lightcone is 
\cite{Dolan:2000ut} 
\be
B(z,\bar{z},\tau,J) = (z \bar{z})^{\frac{\tau}{2}} \left( -\frac{z}{2} \right)^J {}_2F_1(\frac{\tau}{2}+J, \frac{\tau}{2}+J, \tau+2J, z) + \CO(\bar{z}^{\frac{\tau}{2}+1})
\label{eq:LCBlock}
\ee
for an operator with twist $\tau$ and spin $J$. Here,  
$z$ and $\bar{z}$ are defined so $z\bar{z}$ and $(1-z)(1-\bar{z})$ are the standard conformal invariant cross-rations; the important point is that one approaches the lightcone as $\bar{z} \rightarrow 0$. 
In 
$d=2$, the twist $\tau_{\rm min}(n)$ of the lowest-twist multi-stress tensors vanishes, and therefore one can isolate their contribution (assuming there are no other conserved currents in the theory) simply by going to the lightcone.  By contrast, generally 
in $d>2$, the factor $\bar{z}^{\frac{\tau}{2}}$ vanishes at $\bar{z}=0$.  
However, from the gravity side one sees  
that more stress tensors should also be enhanced by more factors of $f_0$. Therefore, by taking a limit of $\bar{z}\rightarrow 0$ with
\be
\label{largef0}
\left[ f_0^n \bar{z}^{\frac{\tau_{\rm min}(n)}{2}} \right]^{\frac{1}{n}} = f_0 \bar{z}^{\frac{d-2}{2}} \textrm{ fixed} \ ,
\ee
we expect to be able to isolate the contributions from the lowest-twist $T^n$s.  
As a consequence, the resulting large $f_0$ limit \eqref{largef0}  
takes the mass of the black hole to infinity in units of the Planck mass, and therefore is a high-temperature limit.  Since the AdS black hole is becoming infinitely large, we can focus on the planar black hole background for simplicity. We will mostly take this approach in this paper, and relegate some preliminary investigations of the spherical black hole case to an appendix.  

To compute the holographic correlator (\ref{eq:2point}), we first solve for the bulk-to-boundary propagator for the light operator $\CO_L$, and then take the bulk point to the boundary. To obtain the bulk-to-boundary propagator, we must solve the bulk equation of motion in the black hole background.  
Rather than solving the bulk field equation in a general black hole background  
exactly, which is not possible analytically, we will 
 extract the boundary two-point function order-by-order in a short distance expansion (in other words, the OPE).  
As the boundary operators approach each other, one might reasonably expect that their two-point function  is sensitive only to bulk physics near the boundary, and therefore order-by-order it depends only on a near-boundary expansion of the bulk metric.  This intuition turns out to be correct for the $T^n$ operator contributions.

We emphasize that our calculation goes beyond the geodesic approximation, where (\ref{eq:2point}) is calculated in the large $\Delta_L$ limit by computing the length of the geodesic between two light operators at the boundary.  In the geodesic approximation, it is clear that the short distance expansion of  (\ref{eq:2point}) depends only on a near-boundary expansion of the bulk metric, since the geodesic itself is constrained to be close to the boundary as the two light operators approach each other. Moreover, as we will demonstrate, the lowest-twist operators in this approximation are controlled by geodesics with large angular momentum that stay close to the boundary, and depend only on the leading expansion of the metric near the boundary.  This fact makes the universality of lowest-twist $T^n$s easy to understand in the geodesic approximation of the two-point function. The full two-point function, however, depends on the bulk everywhere, and in particular one of the boundary conditions that determines the bulk-to-boundary propagator is imposed at the black hole horizon.  
So it is perhaps less trivial than it might seem that the $T^n$ operator contributions are fixed by the behavior of the metric in an expansion near the boundary of AdS.  Nevertheless, we will find that while the boundary condition near the black hole horizon affects the contribution of some operators in the OPE (in particular, it affects the double-trace operators made from two $\CO_L$s), it does not affect the multi-stress tensors. 
A direct argument\footnote{We thank Jared Kaplan for discussions on this point.} is that the $T^n$ contributions are determined by a product of the OPE coefficients $C_{LL T^n}$ and $C_{HH T^n}$ for $T^n$ in the product of $\CO_L \times \CO_L$ and $\CO_H \times \CO_H$, respectively.  The coefficients $C_{LLT^n}$ manifestly cannot depend on the horizon of the heavy state.  But the $C_{LL T^n}$ coefficients are related to the $C_{HHT^n}$ coefficients by the fact that at large $\Delta_L$, the heavy and light states are symmetric under their exchange.\footnote{Here is another intuitive explanation. In perturbation theory, the double-trace operators require computing the full Witten diagram for the four-point function, whereas the contributions from an exchanged operator require computing only the Witten diagram where the bulk-to-boundary propagators are integrated over geodesics \cite{Hijano:2015qja,Hijano:2015zsa,Dyer:2017zef,Castro:2017hpx}.  This result has been demonstrated for single-trace operator exchange in general $d$ and for semiclassical gravity in $d=2$.  Our results imply that it shall also hold for the multiple graviton exchanges in general $d$ at large $C_T$ as well. 
}

In order to prove that the contributions from lowest-twist $T^n$s are universal, we 
find a ``decoupling" limit, performed in suitable variables, of the bulk-to-boundary propagator that keeps track of the lowest-twist $T^n$s and derive a 
reduced  bulk field equation which manifestly depends only on $f_0$ and $\Delta_L$.  
As multi-stress tensors are fully fixed by UV boundary conditions, this reduced equation implies an all-order proof of the universal lowest-twist OPE coefficients. A key technical step is the use of variables, which we denote $w$ and $\rho$ and are defined in (\ref{eq:wDef}) and (\ref{eq:rhoDef}), that let us efficiently separate out different twists. 

Explicitly solving the reduced field equation in general is still non-trivial.   
However, we find significant simplifications in even dimension $d$.  
In particular, the leading-twist part of the bulk-to-boundary propagator in even $d$ admits  a series expansion whose coefficients satisfy a simple recursion relation and 
 initial 
conditions, \eqref{eq:QwuSeries}-\eqref{eq:RecursionBoundaryCondition1}. 
This recursion relation can be solved efficiently to high order, allowing us to numerically resum the lowest-twist contributions and determine the behavior of the two-point function near the lightcone at infinite temperature.  

For generic values of the light operator dimension $\Delta_L$, the series has a finite radius of convergence,  \eqref{eq:convergenceradius}, in the variable $t \sigma^{\frac{d-2}{2d}}$  
where $t$ is time along a ray close to the lightcone and $\sigma$ parameterizes the angle away from the lightcone (see section \ref{sec:LTTherm} for details). 
At $d=2$, we can directly resum the series for any $\Delta_L$, reproducing the previously known result, but for general $d$ we are 
unable to resum and find closed form expressions for generic $\Delta_L$.  
However, numerically we   
can identify the behavior near the edge of the convergence radius, \eqref{edgeform}. 

 At special non-unitary values of $\Delta_L$, specifically at negative integers, we find further 
simplifications: the radius of convergence becomes infinite, and moreover we can resum the series in closed form.  The resulting expressions allow us to consider the asymptotic behavior at large time (in Lorentzian or Euclidean signature) analytically.  For instance, with $\Delta_L=-1$ and even $d$, we find that along a ray at small fixed angle from the lightcone, the two-point function in a planar black hole background evolves as a power-law times 
the following factor: 
\be
 \sum_{k=1}^d \exp \left( A_d f_0^{\frac{1}{d}} \sigma^{\frac{d-2}{2d}}e^{\frac{2\pi i k}{d}} t \right)\ ,
\ee 
where $A_d$ is a prefactor.  
While the above form gives certain hints toward thermalization, it remains unclear in what sense we may define a notion of 
temperature from such an asymptotic expansion.

Finally, while our focus is on the lowest-twist $T^n$s, we also discuss subleading-twist contributions.  
For concreteness, we will consider an example in $d=4$ which is closely related to quasi-topological gravity (QTG) \cite{Myers:2010ru,Myers:2010jv} (also see \cite{Oliva:2010eb}), and 
derive the subleading-twist $T^n$ contributions for the first few values of $n$.

\subsubsection*{Outline}
An outline of the paper is as follows. 
In section \ref{sec2}, we warm up by revisiting the BTZ black hole case, and then performing a leading order (in the OPE) analysis in $d=4$.  
In section \ref{sec3}, after discussing the general gravitational setup and computational scheme,  we adopt a $d=4$ 
example to perform some explicit computations, including a geodesic approximation. 
We will see explicitly from this example that subleading-twist contributions are not universal as they depend on the details of the higher-curvature corrections in the gravity action.   
The section \ref{sec4} focus on the universal lowest-twist. 
We first discuss how to perform a limit that allows a consistent truncation on the bulk field equation. 
This leads to a reduced field equation, \eqref{PROOF}, which determines the lowest-twist OPE coefficients to all orders. 
We then derive a recursion relation from the reduced field equation and investigate resummations, hints of thermal behavior, and radius of convergence. 
The numerical computations suggest a closed form of the lowest-twist convergence radius in even-dimensional CFTs, \eqref{eq:convergenceradius}.  
We conclude with some future problems in section \ref{sec5}.     
In this work, we mostly focus on the planar black hole case but we include some preliminary analysis of the spherical black hole case in appendix \ref{app:spherical}, which includes two conjectures.  
As a reference, some explicit higher-order solutions in the planar black hole case are listed in appendix \ref{app:sol}.

\renewcommand{\theequation}{2.\arabic{equation}}
 \setcounter{equation}{0}
\section{ Leading Order OPE Analysis}  
\label{sec2}

In this section, we will warm up with an analysis of the holographic heavy-light four-point function  at leading order in the OPE.  Only the stress tensor itself shows up at this leading order, but the analysis will illustrate in a simpler setting the basic ideas and methods of the all-orders analysis. We work in the rest frame of the black hole, which is equivalent to using conformal 
transformations to put the heavy operators at the points 0 and $\infty$.  
 
Ideally, we would like to consider all possible matter fields in the bulk with an action constrained only by some general principles,   
 but to make the calculation tractable, we will make simplifying assumptions and take the bulk Euclidean action as 
 \be
\label{phiaction}
&&~~~~ S_{\rm{tot}} =\int d^{d+1}x \sqrt{g} ~\Big( {\cal L}_\phi+ {\cal L}_{\rm grav} \Big) + S_{\rm{bry}}  \ ,  \\
&& {\cal L}_\phi = \frac{1}{2}(\partial \phi)^2 + \frac{1}{2} m^2 \phi^2  \ ,~~  {\cal L}_{\rm grav}=  R+\Lambda+\cdots \ ,
\ee
where $\dots$ are possible higher-derivative curvature corrections and $\phi$ is the bulk field dual to the probe operator ${\cal O}_L$. 
In other words, we consider a simpler case where all other matter fields decouple or can 
be integrated out, and that self-interactions of $\phi$ can be neglected.\footnote{We will comment more on these 
assumptions in the discussion section.}  
At infinite $c$, this approximation is appropriate 
 and was adopted 
in the analysis of the $d=2$ Virasoro vacuum block \cite{Fitzpatrick:2014vua, Fitzpatrick:2015zha}, where the heavy operator $\CO_H$ enters only through the metric that it induces.  
In dimensions $d>2$, we may again consider backgrounds created by heavy operators.  
The boundary term, $S_{\rm{bry}}$, is included to allow a well-defined variational method 
but it plays little role in our analysis as we focus on solving the bulk field equation. 

We shall consider the following general form of a rotationally invariant 
and stationary metric in the Euclidean signature:\footnote{See, for instance, \cite{Wald}.} 
 \be
ds^2 =  
\big(k+r^2 f(r)\big) dt^2 + \frac{dr^2}{k+r^2 h(r)} + r^2 dU_{d-1}^2 \ .
 \label{eq:BHMetric}
 \ee 
The case  $k=1$ with $dU_{d-1}^2 =d\Omega_{d-1}^2$, the metric on $S^{d-1}$, corresponds to a spherical horizon,  
and  $k=0$ with $dU_{d-1}^2 = l^{-2} \sum_{i=1}^{d-1} dx_i^2$ gives a planar horizon.\footnote{We will not discuss a hyperbolic black hole 
in this paper. The boundary metric in this case does not admit a spherically symmetric spatial foliation and thus cannot be associated with a scalar primary as the background state. 
}    
The asymptotic AdS boundary conditions are  \cite{HT85} (see also \cite{Hollands:2005wt}, eq. (92-98))
\be
\label{asyfr}
f(r) = \frac{1}{\ell^2} - \frac{f_0}{r^d} + \dots, \qquad h(r) = \frac{1}{\ell^2} - \frac{h_0}{r^d} + \dots,     
\ee
where $\ell$ is the AdS radius, which we henceforth set to 1, and $\dots$ are higher-order terms in $1/r$.  
As the explicit functional forms of $f(r)$ and $h(r)$ a prior depend sensitively on the gravitational action in \eqref{phiaction}, which generally 
can contain not only the Einstein-Hilbert term but also higher powers of curvature, the 
two-point function (\ref{eq:2point}) thus depends on the details of these functions and become rather complicated.  
However, let us begin by trying to solve for the two-point function in a short-distance expansion,  
where the two light-operators approach each other, i.e. in their OPE limit.  
When these two operators are close, their correlator should depend only on the behavior of the 
bulk fields near the boundary, and in this limit the two-point function can be computed in a large $r$ expansion.

Although in most of this paper we will focus on the  planar black hole limit, 
in this section we will instead start with a spherical 
horizon. Treating the spherical case here will allow us to illustrate the general case, and also it is more simply related to the boundary conformal block decomposition.\footnote{A reason is that there are additional rescalings needed to extract OPE coefficients with a planar black hole, and these rescalings may be naturally figured out if one has gained some experiences by looking at a spherical black hole.}  Thus, we adopt $k=1$ in this section.  Going beyond the leading order in higher dimensions, the spherical black hole story becomes more complicated.

We would like to compute the two-point function \eqref{eq:2point} by solving for the bulk-to-boundary 
propagator in the metric \eqref{eq:BHMetric} with a black hole. The  propagator obeys the bulk scalar field equation 
\be
\label{eq:BulkEOMGen} 
\left( - \nabla^2 + m^2 \right) \Phi 
= 0 \ , 
\ee  
where $m^2= \Delta_L (\Delta_L-d)$ and we identify
\be
 \Phi(r, x_1, x_2) \equiv \< \CO_L(x_1) \phi_L(r,x_2)\>_{\rm BH}  \ .
\ee  
From now on, let us simply write $\Delta_L= \Delta$. 
In Euclidean space the scalar approaches a $\delta$ function at the boundary:\footnote{More precisely, the 
boundary behavior is the $\delta$-function on flat Euclidean space mapped to the appropriate boundary coordinates.}      
\be
\lim_{r \rightarrow \infty} \Phi(r, x_1, x_2) \propto r^{\Delta-d} \delta^{(d)}(x_1-x_2) \ .
\ee
We then take the bulk point to the boundary to obtain 
\be
\label{dictionary}
\<\CO_L(x_1) \CO_L(x_2)\>_{\rm BH} = \lim_{r\rightarrow \infty} r^{\Delta} \Phi(r, x_1, x_2)  \ .
\ee
In pure AdS, $f(r)=g(r)=1$, the bulk-to-boundary propagator has a simple closed form \cite{Witten:1998qj} 
\be
\label{pureAdSsph}
\Phi_{\rm AdS}(r, x_1, x_2) &=  & \left( \frac{1}{2} \frac{ 1}{\sqrt{1+r^2} \cosh t - r \cos \theta } \right)^{\Delta}  \ . 
\ee
We have Wick rotated the difference $ t = t_1-t_2$ in time 
to Euclidean signature in above, and $\theta$ is the angle between $x_1$ and $x_2$.

For general metrics no simple closed form of the bulk-to-boundary propagator exists.  
Instead, one may try to solve for the bulk-to-boundary propagator in the short distance expansion described 
above, with the large $r$ limit together with the OPE limit where the two boundary points $x_1, x_2$ to approach each other: 
\be
r\rightarrow \infty \quad \textrm{ with } \quad \Hat t =r t , \ \Hat \theta= r \theta ~~ \textrm{ fixed }.
\ee
In this limit, the pure AdS propagator \eqref{pureAdSsph} reduces to the simpler form 
\be
\label{definewsph}
\lim_{r\to \infty} \Phi_{\rm AdS} = \left( \frac{r}{w^2} \right)^\Delta, \quad w^2 \equiv 1 + \Hat t^2 + \Hat \theta^2.
\ee
The variable $w$ turns out to be a very convenient coordinate to use.

\subsection{Remarks on BTZ}
\label{sec:BTZ}

To get a sense of what kind of structure we should expect, it will be 
illuminating to revisit the BTZ black hole using the variable $w$ defined above.
In the coordinate system (\ref{eq:BHMetric}) with $k=1$,  
the BTZ metric is 
\be
f(r)=h(r) = 1-\frac{f_0}{r^2} \ ,
\ee and the Schwarzschild radius is $r_+=\sqrt{f_0-1}$.    
In this case, the closed-form result of the bulk-to-boundary propagator is \cite{KeskiVakkuri:1998nw} 
\be
\label{BTZexact}
\Phi_{\rm BTZ}  (r, t, \theta) 
&=& \sum_{n=-\infty}^\infty \left( \frac{-r_+^2/2}{\sqrt{r^2-r_+^2} \cos(r_+ t) - r \cosh\big(r_+ (\theta+2\pi n)\big)} \right)^\Delta .
\ee
The pure AdS case corresponds to taking $r_+=i$ and keeping only the $n=0$ term.  
This $n=0$ in general is the contribution from the modes that correspond to the boundary stress tensor and 
its products with itself \cite{Fitzpatrick:2014vua}, whereas the $n\ne 0$ terms correspond to the double-trace modes, 
\be
[ \CO_L^2]_{n,\ell} \sim \CO_L (\partial)^{2n} \partial^{\mu_1} \dots \partial^{\mu_\ell} \CO_L  \ ,
\ee
which are made from two probe operators.  
Now, making a change of variables from $(r, t, \theta)$ to $(r, w, \hat \theta)$, the expansion of the $n=0$ term of the propagator \eqref{BTZexact} can be written as  
\be
\label{BTZn0}
\frac{\Phi_{{\rm BTZ }, n=0}}{\Phi_{\rm AdS}} = 1 + \frac{f_0 \Delta}{r^2} \frac{w^4 +4 w^2 -8 -2 \th^2 (2+w^2)}{12 w^2} + \sum_{m=2}^\infty \frac{p_m(w,\th)}{r^{2m} w^{2m}}  \ .
\ee
In above, we have explicitly written out the first two terms in the large $r$ expansion, and the others 
are all of the form indicated where $p_m(w,\th)$ is a polynomial in $w,\th$ of maximum order $w^{4m}$ in $w$ and $\th^{2m}$ in $\th$. 
To obtain the boundary-boundary correlator, we next rewrite $w,\th$ in terms of $t,\theta$ and take 
the $r\rightarrow \infty$ limit to read off the coefficient of $r^{-\Delta}$: 
\be
\lim_{r \rightarrow \infty} r^{\Delta} \Phi_{{\rm BTZ }, n=0} = \frac{1}{(t^2+\theta^2)^\Delta} \left( 1 + f_0 \Delta \frac{(t^2+\theta^2)- 2 \theta^2}{12} + \dots \right).
\ee

One may also consider the $n\ne 0$ terms, although we will not focus on double-trace modes.  
In the large $r$ limit with $w, \th$ fixed,  
\be
\label{BTZn1}
\frac{\Phi_{{\rm BTZ}, n\ne 0}}{\Phi_{\rm AdS}} 
=  \left( \frac{ w}{r} \   \frac{r_+}{2 \sinh(\pi n r_+)} \right)^{2\Delta} \left( 1 - \frac{r_+ \th \Delta \coth (\pi n r_+)}{r} + \sum_{s=2}^\infty \frac{q_s(w,\th)}{r^{s} w^{s}} \right) \ .
\ee 
We have again only shown the first two terms in the large $r$ expansion, which is of the general form indicated with $q_s$ a polynomial in $w$ and $\th$. 

It turns out that much of the polynomial structure 
that we obtained in the BTZ black hole case largely holds in higher dimensions as well.
In general $d$, the bulk field equation (\ref{eq:BulkEOMGen}) allows only $r^{-\Delta}$ and $r^{\Delta-d}$ as leading terms 
in a large  $r$ series expansion; as mentioned, the coefficient of $r^{\Delta-d}$ is fixed to be a $\delta$-function, and the 
coefficient of 
$r^{-\Delta}$  
is the two-point function we want.  
As we consider the classical bulk-to-boundary propagator on a fixed background metric, 
the only operators in the conformal block decomposition of the two-point function 
are powers of stress-tensor and double-trace operators. 
This fact not only immediately dictates the only allowed powers of $t$ and $\theta$ in the two-point function, but also constrains the coefficients of 
all the other terms in the $1/r$ series expansion of $\Phi$ since the bulk field equation relates higher-order terms to derivatives of lower-order terms.

\subsection{Leading Order OPE in $d=4$}
\label{sec:LeadingD4Analysis}

As a warm-up, let us here solve for the first correction term of the large $r$ expansion, taking $d=4$ for concreteness and simplicity. By ``leading order'', we mean leading in the OPE, which corresponds to the leading $1/r$ correction to the scalar field $\Phi$.\footnote{This does not mean the gravity background has no higher-order curvature corrections.   
On the other hand, if the background is simply $f=h=1-{f_0\over r^4}$, there are still higher-order corrections to $\Phi$,  corresponding to higher-order OPE computed in the AdS-Schwarzschild geometry.} 
We write
\be
\label{G0ans}
\Phi &=& 
\Phi_{\rm{AdS}} \left( 1 + \frac{G_0(w,\th)}{r^4} + \dots \right) 
\ee
in the limit of large $r$ with $w, \th$ fixed.  
By power counting, the only operator that can contribute to $G_0$ is a single stress tensor. 

In the BTZ case above, the coefficient of $r^{-2}$ was $w^{-2}$ times a polynomial in $w,\th$ -- see \eqref{BTZn0}. 
A similar polynomial form
turns out to be true of $G_0$ in any even $d$, and thus an efficient way to solve for $G_0$ might be simply 
taking an Ansatz with sufficiently high-order polynomial.  
However, it will be more illustrative, and  more representative of general $d$, to start with the less restrictive condition 
that $G_0$ be a polynomial in $\th$ of at most $\CO(\th^2)$,
\be
\label{G0th2}
G_0(w,\th) = a^{(0)}(w) + a^{(2)}(w) \th^2 \ ,
\ee
as one might expect since the stress tensor is spin 2 and $\th$ is an angular variable.  

Taking $d=4$ 
and substituting \eqref{G0ans}, \eqref{G0th2} into the bulk field equation, 
one obtains ordinary differential 
equations for $a^{(0)}(w)$ and $a^{(2)}(w)$ that can be solved analytically.\footnote{One solves the scalar field equation order-by-order in $\th$. 
The equation associated with the highest power of $\th$ involves $a^{(2)}(w)$ only, and the next order equation contains both $a^{(0)}(w)$ and $a^{(2)}(w)$.}
We find   
\be
a^{(2)}(w) = \frac{ \left(10 w^4-15 w^2+6\right)\Delta  f_0}{30
   \left(w^2-1\right)^3 w^2}+\frac{c_1 w^8}{256 \left(w^2-1\right)^3}+\frac{c_2 \left(\frac{w^4}{\Delta -2}-\frac{2 w^2}{\Delta -3}+\frac{1}{\Delta
   -4}\right) w^{2 \Delta }}{512
   \left(w^2-1\right)^3} \ .
   \ee
   The integration constant $c_2$ must vanish since otherwise it would modify the coefficient of $r^{\Delta-4}$ in 
$\Phi$ at large $r$ with 
$t,  
\theta$ fixed.  
The integration constant $c_1$ is fixed by demanding regularity at $w=1$, since 
otherwise there would be an unphysical singularity in $\Phi$ with the  bulk and boundary operator at finite separation. 
The result is    
\be
a^{(2)}(w) = -  \frac{ w^4+3 w^2+6 }{30 w^2}  \Delta  f_0 \ .
\ee
The same constraints fix the integration constants for $a^{(0)}(w)$. We have 
\be
\label{a0w}
a^{(0)}(w)&=& {1\over 120} \Big[ \frac{ (4 \Delta -10) f_0-(\Delta -4) h_0 }{\Delta -2} \Delta w^4+\frac{2(4 \Delta -7) f_0-2(\Delta -4) h_0}{\Delta -1}  \Delta  w^2 \nn\\
&&~~~~~~~ +6 \big((2 \Delta +3) f_0+(2 \Delta -3) h_0\big) -\frac{24\left(f_0+h_0\right)}{w^2}  \Delta \Big]  \ .
\ee 

There are already two immediate reasons to impose condition $h_0=f_0$. First is to notice that the poles 
at integer $\Delta$ disappear only if the background satisfies $h_0=f_0$.  As there is no double-trace modes 
at this leading order, the condition $h_0=f_0$ is the only way to remove these poles.  
At higher orders, there are double-trace modes and we shall require that the total contribution is regular at integer $\Delta$.  
The second reason is even simpler: $G_0$ should vanish when $\Delta=0$, and this requires $h_0=f_0$. 
In below, we will provide a perhaps more interesting reason why $h_0=f_0$ is a required condition based on conformal invariance. 
That is, the conformal block decomposition in the boundary limit requires $h_0=f_0$.

Assuming $h_0 \neq f_0$, the final contribution is obtained by taking the $r \rightarrow \infty$ limit with $t,\theta$ fixed:  
   \be
\label{finalcontri}
 \lim_{r \rightarrow \infty}  {G_0(w,\th)\over r^4} = {\Delta\over 120 (\Delta -2)}  \left(t^2+\theta ^2\right) \Big((4 \Delta -10) f_0
   t^2-2 f_0 \theta ^2-(\Delta -4) h_0 \left(t^2+\theta ^2\right)\Big)\ .
   \label{eq:g2LeadingTerm}
   \ee
To process \eqref{finalcontri} into a formula for the OPE coefficients of the stress tensor, we compare to the conformal block for the stress tensor.  
In conventional $z,\bar{z}$ variables, the leading terms of the block are \cite{Dolan:2000ut,Dolan:2003hv}\footnote{See \eqref{B}. We have divided out the Generalized Free Field (GFF) theory factor, $(z\bar{z})^{-\Delta}$.}   
\be
B(z, \bar z, \tau=d-2, J=2) = (z \bar{z})^{\frac{d-2}{2}} \left( z^2 + 2(1-\frac{2}{d}) z \bar{z} + \bar{z}^2+ \dots \right)\ ,
\label{eq:BlockLeadingTerm} 
\ee
where twist $\tau=\Delta_T-J$ with $\Delta_T$ the dimension of internal stress-tensor 
operators and $\dots$ above denote higher orders in $z,\bar{z}$.   
The relations between coordinates $z,\bar{z}$ and $t,\theta$ are
\be
(1-z) = e^{t+i \theta} \ , ~~~ (1-\bar{z}) = e^{t-i \theta}\ .
\label{eq:zzbarTOttheta}
\ee
Comparing the RHS of (\ref{eq:g2LeadingTerm}) with the RHS of (\ref{eq:BlockLeadingTerm}) in $d=4$, we find that  
the matching 
between the correlator and the conformal block 
is possible only if 
\be
\label{eq:f0h0}
h_0 = f_0\ .
\ee
Evidently, enough symmetry is left after inserting the heavy operators to impose this constraint on the background metrics that they source.  

We can now read off the coefficient of the stress-tensor block:\footnote{In the following, we adopt the notation $c_{\rm OPE} (\Delta_T, J)$ instead of  $c_{\rm OPE} (\tau, J)$, which was used in \eqref{f0sum} to emphasize the lowest-twist. 
We shall normalize $c_{\rm OPE}$ as the coefficient of $(z \bar{z})^{\frac{d-2}{2}} z^J$ in the leading term of the conformal block \eqref{eq:BlockLeadingTerm}.  
This differs by a factor of $(-\frac{1}{2})^J$ from the convention used in Dolan and Osborn \cite{Dolan:2000ut}.  
In particular, $c_{\rm OPE}(d,2)|_{\rm here} 
= \frac{1}{4}  c_{\rm OPE}(d,2)|_{\rm there} 
= \frac{\Delta_L \Delta_H d^2}{4 C_T (d-1)^2}$.   
On the other hand, our expression for $c_{\rm OPE} (d,2)$ in general $d$, given later in \eqref{eq:Pd2OPE},  
agrees with eq. (3.31) in \cite{Kulaxizi:2018dxo} after translating conventions $\lambda_{LLT} \lambda_{HHT}=
 4 c_{\rm OPE}(d,2)|_{\rm here}$ and   
$\mu= f_0|_{\rm here}$.
}
\be
c_{\rm{OPE}}(4,2) \equiv c_{LLT} c_{HHT} = \frac{\Delta}{120} f_0 \ .
\ee
As indicated, this coefficient $c_{\rm{OPE}}(4,2)$ is the product of the OPE coefficients for $ \CO_L \CO_L \sim T_{\m\n}$ 
and $\CO_H \CO_H \sim T_{\m\n}$.\footnote{The normalization of the stress tensor here is a somewhat less common one 
where the central charge $C_T$ is absorbed into the normalization of the operator $T_{\m\n}$.}  
A  symmetry argument gives\footnote{To make the argument precise, one can formulate the expansion as an expansion in powers of $\Delta/C_T$ and $\Delta_H/C_T$, so that it is symmetric between $\Delta \leftrightarrow \Delta_H$.}
\be
c_{HHT}= \lim_{\Delta \to \Delta_H }c_{\rm{OPE}}(4,2)^{1\over 2} \ . 
\ee 

The leading-order analysis we have just done above illustrates several basic ideas we will implement in the rest of the paper. 
The analysis however will become more technical as we go to higher orders, 
but the underlining techniques are largely the same.  
The exception will be the geodesic analysis, which is a somewhat orthogonal technique that makes some aspects of the physics more transparent.

As we go to higher orders, 
we will find contributions of more and more operators, and their contributions will depend not only on $f_0$ but also on the details of functions $f(r)$ and $h(r)$.  
Of the two types of contributions, multi-stress-tensor and double-trace operators, we mainly restrict our attention to the multi-stress-tensor contributions, denoted as $T^n$s.  
As we will see, although these $T^n$ contributions depend sensitively on the forms of $f(r), h(r)$, and thus essentially require knowledge of an infinite number of parameters, there is a rather 
special class of lowest-twist $T^n$ primary operators whose contributions turn out to be protected: they 
depend universally only on $f_0$ and the weight $\Delta$ of the probe operator.  

The lowest-twist $T^n$ primaries at each $n$ are mapped to the lowest-twist product of $n$ stress tensors.  
As twist is dimension minus spin, the lowest-twist is achieved by leaving all indices uncontracted, e.g. 
\be
[T^n]_{\tau_{\rm min}} \sim T^{\mu_1 \nu_1} \dots T^{\mu_n \nu_n} \ .
\label{eq:LTOps}
\ee
It is possible to create additional primary operators with the same $n$ factors of $T^{\mu\nu}$ and the 
same twist by sprinkling factors of $\partial^\mu$ with uncontracted indices among the $T$s in $[T^n]$.   
However, in the planar limit, 
adding powers of derivatives $\partial^\mu$ causes the contributions to scale to zero, in contrast 
with adding factors of $T^{\mu\nu}$ which bring along extra 
compensating powers of the temperature. Thus, the only lowest-twist $T^n$ 
primaries that contribute in the planar limit are those of the form in (\ref{eq:LTOps}).   
We will therefore be mostly interested in the case of planar black holes in this work, which correspond to essentially black holes 
in the high-temperature limit where one simultaneously scales the CFT spacetime to compensate. 

The spherical (i.e. finite temperature) black hole case will be left to 
appendix \ref{app:spherical}, where we will argue that the contributions 
of these $T^n$ primary operators again universally depend only on $f_0$ and $\Delta$.

\renewcommand{\theequation}{3.\arabic{equation}}
 \setcounter{equation}{0}
\section{Higher-Derivative Gravity and Conformal Blocks}
\label{sec3}

The central theme of this work is to probe the conformal blocks of CFTs 
in the holographic framework by including higher-derivative curvatures on the gravity side 
and search for universality.
A general argument for the universality at lowest-twist will be given in the next section. 
However, it will be instructive to first adopt a concrete example to explicitly solve for the bulk-to-boundary 
propagator using a near-boundary expansion and obtain some  higher-order OPE coefficients.
In this section, we shall start with our general setup and then consider a $d=4$ 
example, which is closely related to quasi-topological gravity \cite{Myers:2010ru,Myers:2010jv} (see also \cite{Oliva:2010eb}). 
Aside from being a warm-up for the general approach presented in the next section, this specific example will allow us to see directly that the sub-leading twist OPE coefficients are 
non-universal, i.e. they are contaminated by other generally model- and coupling-dependent parameters. 

\subsection{General  Setup}

\subsubsection{Gravitational Action}

Symbolically, we may write the most general higher-derivative gravity action as 
\be
\label{mostgeneralS}
S_{\rm grav}= \int d^{d+1}x \sqrt{g}  \left(R+\Lambda+\sum_i  \alpha_i {\cal O} (R^2)+\sum_j \beta_j {\cal O} (R^3) \ ,
+\sum_k \gamma_k{\cal O} (R^4)+\cdots \right) 
\ee  
where ${\cal O} (R^{\#})$ denotes all possible Lorentz invariants constructed out of the Riemman curvature 
tensor $R_{\mu\nu\lambda\rho}$ and metric $g_{\mu\nu}$ with powers $\#$ fixed.  Indices $i,j,k,...$ represent the numbers of independent invariants.  
For concreteness, we here focus on a planar black hole and 
consider a static, spherically symmetric metric
\be
\label{metricFH}
ds^2= r^2 f(r) dt^2  + {dr^2 \over r^2 h(r) } + r^2 \sum_{i=1}^{d-1}dx_i^2\ ,
\ee 
where  the functions $f(r)$ and $h(r)$ (black hole solutions) depend sensitively on the 
coefficients in the gravity action \eqref{mostgeneralS}.    
By turning off higher-derivative corrections, the theory reduces 
to Einstein gravity with a negative cosmological constant.  

Here we are interested in solving the scalar field equation 
\be
\label{EoM}
(\Box - m^2) \Phi =0 \ , ~~~~~~m^2=\Delta\Big(\Delta-d\Big)  \ , 
\ee 
in the background \eqref{metricFH}, subject to the $\delta$-function boundary condition.
As the background asymptotes to AdS, we may treat the metric as being AdS plus a perturbation.   
Parametrizing the coordinates using variables $(t,r, \rho)$ where 
\be
\rho^2=\sum_{i=1}^{d-1} x_i^2 \ ,
\label{eq:rhoDef}
\ee 
the field equation  \eqref{EoM} can be written as 
\be
\label{PDE}
 &&\Big[ {\del^2_t\over r^2 f} +{r^2 h \over 2} \Big( {f'\over f}+{h'\over h}  
+{10\over r} +  2  \del_r\Big)\del_r +{1\over r^2} \big({2\over \rho}+\del_\rho\big)\del_\rho-  (\Delta-4) \Delta \nn\\
&&~~~~~~~~~~~~~~~~~~~~~~~~~~~~~~~~~~~~~~~~~~~ + (d-4)\Big(\Delta+ {1\over \rho} {\del_\rho \over   r^2}+ h r\del_r \Big) \Big] \Phi = 0  \ .
\ee  We have factored out a $(d-4)$ so that the last 
piece above can be dropped conveniently in the $d=4$ example considered later. 

\subsubsection{Change of Variables}

Identifying suitable variables is normally an important step in analyzing PDEs, such as \eqref{PDE}.  
To better analyze \eqref{PDE}, we find it useful to first define re-scaled $t$ and $\rho$ as 
\be
(\Hat t, \Hat \rho)= r (t, \rho) \ .
\ee 
Starting with the canonical variables $(r, t, \rho)$, we next consider the following change of variables:
\be
\label{holyw}
(r, t, \rho) \to (r, \Hat t, \Hat \rho)  \to  (r, w , \Hat \rho) \ , 
\ee  
where we introduce  
\be
w^2= 1+ \Hat t^2+\Hat \rho^2 \ .
\label{eq:wDef}
\ee
This variable $w$ is perhaps naturally suggested already by the free-propagator: 
\be
\label{freesol}
\Phi_{\rm{AdS}}(r, t, \rho)=\Big({r\over  1+ r^2 (t^2+  \rho^2)}\Big)^\Delta\equiv \Big({r\over  w^2}\Big)^\Delta \ ,
\ee  which solves the field equation \eqref{PDE} in pure AdS.   

Writing
\be
 \Phi (r, w , \Hat \rho)&=& \Phi_{\rm{AdS}}  G(r, w , \Hat \rho) \ ,
\ee  
the field equation \eqref{PDE} in terms of the new variables $(r, w , \Hat \rho)$ can be written as
\be 
\label{PDEwy}
&&\Big[\del^2_r +C_1\del^2_w+ C_2 \del^2_{\y}+C_3\del_r\del_{w}+ C_4\del_r\del_{\y}+C_5\del_w\del_{\y}\nn\\
&&~~+C_6\del_r+ C_7 \del_w+C_8 \del_{\y}+C_9\Big] G=0 \ ,
\ee
where the $C_i$ coefficients are
\begin{eqnarray}
&& C_1 = {f \big(\Hat \rho^2 + (w^2-1)^2 h\big)+w^2 - \Hat \rho^2-1  \over r^2 w^2 f h} \ ,\\
&& C_2 = {1+ h {\Hat \rho}^2 \over r^2 h}\  , \\
&& C_3 = {2 \over r w} (w^2-1) \ ,\\
&& C_4 =  {2 \Hat \rho\over r } \ , \\
&& C_5 =  {2 \Hat \rho\over r^2 w h} \big(1+(w^2-1) h\big)\ ,
\end{eqnarray}
\begin{eqnarray}
&& C_6 = {f'\over 2f}+{h'\over 2h}  +{w^2 \big(10 - 4 \Delta\big)+8 \Delta \over2 r w^2}  +{(d-4)\over r}\ ,  \\
&& C_7 = \Big( {h'f+hf' \over 2 r w f h}- {w^2 (2 \Delta-5) -4\Delta-1\over r^2 w^3}\Big)(w^2-1)\nn\\
&&~~~~~~~~+{3 w^2 - {\Hat \rho}^2 (1 + 4 \Delta)\over r^2 w^3  h} + { 1 + {\Hat \rho}^2 +  4 (1 - w^2 + {\Hat \rho}^2) \Delta  \over r^2 w^3 f h} +(d-4) {C_5\over 2 \y}
\ , 
\end{eqnarray}
\begin{eqnarray}
&& C_8 ={ 2 (w^2 - 2 {\Hat \rho}^2 \Delta) + \Hat \rho^2 \big(w^2 (5 - 2 \Delta) + 4 \Delta\big) h \over r^2 w^2 \Hat \rho h}+{\Hat \rho(h'f+hf')\over  2 r f h} +(d-4)  {C_2\over \y}  
 \ ,
\end{eqnarray}
\begin{eqnarray}
&&{C_9}={\Delta\over w^2} \Big[{( w^2-2)^2 \Delta +4 (1+w^2-w^4)   \over r^2 w^2}\nn\\
&&~~~~~~~~~~~~~~ +{  4 {\Hat \rho}^2 (\Delta+1)  - w^4 ( \Delta-4)- 6w^2\over r^2 w^2 h} +{2w^2 (1 + 2 \Delta)-4 (1 + {\Hat \rho}^2) (1 + \Delta) \over r^2 w^2 f h}\nn\\
&&~~~~~~~~~~~~~~ - {(w^2-2) (h'f+hf') \over 2 r f h}  - (d-4){(w^2-2)(h-1)\over r^2 h} \Big]  \ .
\end{eqnarray}

Admittedly, the field equation written in the new variables $(r, w, \y)$ looks more complicated when compared with \eqref{PDE}. 
The structure of perturbative solutions, discussed below, however becomes simpler to analyze. 
We will also see that adopting the variable $w$ turns out to be a crucial step toward finding a general proof of the universal lowest-twist coefficients.

\subsubsection{Near-boundary Structure}
\label{nbs}

We consider solving the field equation \eqref{PDEwy} in a large $r$ expansion, which corresponds to a short-distance 
expansion of the light operators.  
One may first formally write    
\be
\label{generalf}
&&f(r)= 1- {1\over r^d} \sum_{i=0,1,2,..} {f_i \over r^i} 
= 1- {f_0\over r^{d}}-{f_1\over r^{d+1}}-{f_2\over r^{d+2}}- \dots \ , \\
\label{generalh}
&&h(r)= 1- {1\over r^d} \sum_{i=0,1,2,..} {h_i \over r^i} 
= 1- {h_0\over r^{d}}-{h_1\over r^{d+1}}-{h_2\over r^{d+2}}- \dots \ ,
\ee  
as the asymptotic expansions.
The general gravity action \eqref{mostgeneralS} a priori allows solutions with $f_i\neq h_i$ but 
exploring black hole solutions in the context of higher-derivative gravity is not the main focus of the 
 present paper.\footnote{For a recent review on higher-order gravities and references see, e.g., \cite{Bueno:2016ypa}.}    
Indeed, as advertised and will be proven, our main result that the 
lowest-twist coefficients are protected does not rely on the details of black hole solutions. 

In general, there are both stress-tensor and double-trace solutions.  
The double-trace modes can only be determined by an interior boundary condition and 
a near-boundary analysis becomes invalid in such an IR region.    
While these modes are entangled at integer $\Delta$, there is a clean separation between them at non-integer $\Delta$.
We will thus mainly focus on the $T^n$ conformal blocks with non-integer $\Delta$. 
Some formal expressions of the double-trace solutions will still be included below. 

Conformal symmetry, and in particular requiring that the boundary correlators can be decomposed into conformal blocks of physical states, imposes constraints on the gravitational background.  
A factor ${f_i\over r^{d+i}}$ or ${h_i\over r^{d+i}}$ in the background 
generally induces a corresponding order solution, $G\sim {G_i (w, \y) \over r^{d+i}}$, in the scalar perturbative solution.   
In the large $r$ limit, it leads to a finite contribution,  
$\lim_{r\to \infty} {G_i (w, \y) \over r^{d+i}} \sim  (t^2+\rho^2)^s  \rho^{d+i-2s}$ for some integer $s$.  
Changing variables to $z, \bar z$, it contributes to the boundary correlator a term 
of order ${\cal O}(z \bar z)^{d+i}$.\footnote{The details of changing variables and relevant rescalings in the planar black hole case are given in \eqref{transf} and \eqref{rescaling}, respectively.} 
The $T^n$ conformal blocks in the high-temperature limit however only allow certain restricted powers: ${\cal O}(z \bar z)^{\a}$, $\a=d,2d,3d,...$.
Therefore, instead of the arbitrary \eqref{generalf}, \eqref{generalh}, we shall start with  
\be
\label{ourf}
&&f(r)= 1- {f_0\over r^{d}}-{f_d\over r^{2d}}-{f_{2d}\over r^{3d}}- \dots \ , \\
\label{ourh}
&&h(r)= 1- {h_0\over r^{d}}-{h_d\over r^{2d}}-{h_{2d}\over r^{3d}}- \dots \ .
\ee  
Recall that the leading-order analysis leads to the condition $h_0=f_0$. 
We will find that higher-order factors $f_i, h_i$ with $(i>0)$ generally do not have such a restriction.     

We next discuss the general structure of the scalar field solution.
Denote $G^{T}$ and $G^{\phi}$ as multi-stress-tensor and double-trace contributions, respectively. 
We write the general solution to the field equation \eqref{PDEwy} as 
\be
 \Phi = \Phi_{\rm{AdS}} G\ , ~~~G = 1+G^{\rm{T}}(r, w, \y)+ G^{\rm{\phi}}(r, w, \y)\ , 
\ee 
with
\be
G^{T} ={1\over r^d} \sum_{i=0,d,2d,...} {G^T_{i}(w, \y)\over r^i} \ , ~~~G^{\phi} = \big({w\over r}\big)^{2\Delta} \sum_{i=0,1,2,...} {G^\phi_{i}(w, \y)\over r^{i}} \ .
\label{eq:Gexpansion2}
\ee  
From a search of the general pattern of perturbative solutions 
and also suggested by the BTZ analysis in the previous section, we  can 
identify the structures of  $G^T_{i}(w, \y)$ and $G^\phi_{i}(w, \y)$.  
\\

{\rm $Stress~Tensor$:}
\be
\label{STsol} 
&&G^T_{i}= \sum_{j=0,2,4,..}^{2(1+{i\over d})} \a^{(j)}_{i}(w)  \y^j \ , ~~~~ i=0,d,2d,... 
\ee 
For instance, in $d=4$,
\be
&&G^T_{0}=\a^{(0)}_0(w)+ \a^{(2)}_0(w)\y^2\ , \\  
&&G^T_{4}=\a^{(0)}_4(w)+ \a^{(2)}_4(w)\y^2+ \a^{(4)}_4(w)\y^4 \ , \\   
&&G^T_{8}=\a^{(0)}_8(w)+ \a^{(2)}_8(w)\y^2+ \a^{(4)}_8(w)\y^4+\a^{(6)}_8(w)\y^6 \ .    
\ee   
Only the leading-power in $w$ in the solutions $\a^{(j)}_{i}(w)$ survive at large $r$ and they determine the OPE coefficients. 

To understand why the powers of $\y$ truncate in the above manner is a bit more 
involved, but the principal requirement is the consistency with the conformal block 
decomposition.   
Let us try to illustrate this point through a simple example. 
Say, in the $d=4$ ${G^T_{0}}$ solution, one considers an 
$\a^{(s)}_0(w)\y^s$ term for some integer s. In the boundary limit, the relevant pieces are $\sim  w^{4-s} \y^s$.
Considering now an infinite series, one has 
\be
\label{A}
\lim_{r\to \infty} {1\over r^4}\sum_{s=-\infty}^{\infty}  c_s  w^{4-s} \y^s 
=
\sum_{s=-\infty}^{\infty}c_s  {(-1)^{s\over 2}\over 2^s} {(z-\bar z)^s \over (z \bar z)^{(s-4)\over 2}}   \ , 
\ee  
where the constant coefficients $c_s$ are proportional to $f_0$.   
On the other hand, the stress-tensor conformal block at this level has the structure
\be
\label{B}
c_{\rm OPE} ~ z \bar z (z^2 + z \bar z + \bar z^2) \ ,
\ee  
where $c_{\rm OPE}$ is the OPE coefficient, corresponding to a single stress-tensor exchange in the present case.  
Equating \eqref{A} with \eqref{B}, the uniqe solution is 
\be
c_0=3 c_{\rm OPE} \ , ~~~ c_2= -4  c_{\rm OPE} \ , 
\ee 
and $c_s=0$ for $s >2$, which explains the trucation in $\y$. 
One can go further and look at higher orders to identify the general pattern \eqref{STsol}.    

We next discuss the general structure of the functional coefficients $\a^{(j)}_i(w)$.
Plugging \eqref{STsol} into the field equation and solve for $\a^{(j)}_{i}(w)$, the solutions can be written as
\be
\label{aw}
A(w)+c_1 B(w)+ c_2 w^{2\Delta} C(w) \ .
\ee  
In general, we must set $c_2=0$ to preserve the $\delta$-function boundary condition.
Moreover, we find generally that the remaining integration constant $c_1$ can be fixed by requiring regularity at $w=1$.  
The resulting solutions $\a^{(j)}_{i}(w)$ admit polynomial forms.

The resulting $\a^{(j)}_{i}(w)$ may have poles at integer 
$\Delta$. These poles, if exist, are expected to be canceled by including the double-trace modes because the full scalar solution should be regular at any $\Delta$.
\\ 

{\rm $Double~Trace$:} 
\be
\label{DTsol}
&&G^\phi_{i}= \sum_{j=0,1,2,..}^{i} \beta^{(i-2j)}_{i} w^{2j}  \y^{i-2j}  \ , ~~~(\beta^{(j)}_{i} =0  ~~{\rm if~} i< 2j )\ .
\ee
For instance, 
\be
G^\phi_{0}=\beta^{(0)}_0\ , ~~ G^\phi_{1}=\beta^{(1)}_1 \y \ , ~~ G^\phi_{2}=\beta^{(0)}_2 w^2+ \beta^{(2)}_2\y^2 \ , ~~ G^\phi_{3}=\beta^{(1)}_3 w^2 \y+\beta^{(3)}_3 \y^3 \ .
\ee 
The constant coefficients $\beta^{(j)}_i$ should be interpreted as integration constants. 
One may partially determine $\beta^{(j)}_i$ by requiring that integer-$\Delta$ poles from the 
stress-tensor parts should be removed such that the total scalar solution is smooth in any $\Delta$. 
This  condition however still leaves certain ambiguity in $\beta^{(j)}_i$ as one is free to shift $\beta^{(j)}_i$ with some function $k(\Delta)$.
We expect that an interior boundary condition, such as regularity of the bulk-to-boundary propagator at the black hole horizon, has to be imposed to fully determine $\beta^{(j)}_i$.

\subsection{Example}
\label{example}

Although the derivation of the universal lowest-twist will not rely on a specific model 
or spacetime dimensionality, it is useful to consider a concrete higher-derivative gravity example.
Here we shall perform some explicit computations in $d=4$ with background
\be
\label{bg1}
f(r) = 1 - \frac{f_0}{r^4} - \frac{f_4}{r^8} - \frac{f_8}{r^{12}} - \dots \ , ~~ h(r) = 1 - \frac{h_0}{r^4} - \frac{h_4}{r^8} - \frac{h_8}{r^{12}} - \dots \ .
\ee   
Our focus is not searching for gravity actions that lead to the above solutions, but let us mention  
a special higher-derivative gravity model that gives \eqref{bg1}, but with a stronger condition, $f(r)=h(r)$.

\subsubsection*{Quasi-Topological Gravity}

The quasi-topological gravity \cite{Myers:2010ru,Myers:2010jv} (also see \cite{Oliva:2010eb}) 
contains up to curvature-cubed interactions with the specific combinations of Riemann tensors that do not lead to additional 
states around flat space.  The gravity action reads  
\be
\label{5Daction}
S&=& -\frac{1}{16 \pi G} \int d^5 x \sqrt{g} \Big(R+\frac{12}{l^2}+\lambda {\cal L}^{(\lambda)}+\mu  {\cal L}^{(\mu)}\Big)\ , \\
{\cal L}^{(\lambda)}&=& \frac{l^2}{2} \big(R^2_{\mu\nu\rho\sigma}- 4 R^2_{\mu\nu} + R^2\big)\ , \\
{\cal L}^{(\mu)}&=& \frac{7 l^4 }{8}  \Big[ R_{\mu\nu}{}^{\rho\sigma} R_{\rho\sigma}{}^{\alpha\beta}
R_{\alpha\beta}{}^{\mu\nu} +\frac{1}{14}\Big(21 R^2_{\mu\nu\rho\sigma}R
-120 R_{\mu\nu\rho\sigma}R^{\mu\nu\rho}{}_{\alpha}R^{\sigma\alpha}\nn\\
&&~~~~~~~ + 144\,R_{\mu\nu\rho\sigma} R^{\mu\rho}R^{\nu\sigma}
+128\,R_\mu{}^{\nu}R_\nu{}^{\rho}R_\rho{}^{\mu} - 108
R_\mu{}^{\nu}R_\nu{}^{\mu}R +11\,R^3\Big)\Big] \ ,  
\ee  
where $\lambda$ and $\mu$ denote the couplings. 
We have kept the AdS radius in the action \eqref{5Daction} but will set $l=1$ in below.
The Einstein gravity is supplemented not only by the Gauss-Bonnet 
term, ${\cal L}^{(\lambda)}$, which contains curvature-squared interactions, but also 
by curvature-cubed interactions ${\cal L}^{(\mu)}$.  
The linearized field equation of the theory turns out to be 
second-order in AdS$_5$ background and matches the linearized equations of Einstein gravity. 
These $\CO(R^3)$ terms allow one to expand the class of dual CFTs without 
supersymmetry and also allow one to explore the stress-tensor three-point function 
with the full range of parameters. 

This special gravitational theory admits a single-function solution.
Focusing on the planar black hole, we write   
\be
\label{5Dg}
ds^2 &=& N^2(r)\Big(r^2 f(r)\Big) dt^2 + \frac{dr^2}{r^2 f(r)} + r^2 \sum^3_{i=1}d x_i^2 \ .
\ee 
Evaluating the action with this metric leads to a $\delta N$ equation of motion,
\be
1- f(r) + \lambda f^2(r) + \mu f^3(r) &=& \frac{r_0^4}{r^4} \ ,
\ee for some constant $r_0$. 
As the $\delta f$ equation of motion is satisfied for any constant value of $N(r)$, one may simply take 
$N= 1$ for simplicity.
We would like to dial the parameters $\mu$ and $\lambda$ 
which represent new couplings in a small region around the Einstein gravity limit to 
fill out a three-dimensional parameter space:  
one direction is controlled by $\ell_p$ which connects different Einstein theories, and the other 
two directions moving tangentially to this line of 
theories are controlled by the couplings $\mu$ and $\lambda$.  
To quadratic-order in $\lambda, \mu$,  
\be
f(r) &=& f_0(r) + f^2_0(r) \Big(\lambda + \mu f_0(r) \Big) \Big(1 + 2 \lambda f_0(r) 
+ 3\mu f_0^2(r)\Big)  + \CO(\lambda^3, \mu^3) \ , 
\ee 
where $f_0(r) = 1- \frac{r_0^4}{r^4}$.  
The deviation of the metric from pure AdS 
may be treated perturbatively in $r_0\over r$ at large $r$.   
Up to the order $r^{-8}$, 
\be
f(r) &=& 1+(\lambda+\mu)(1+2\lambda+3\mu) -  \left( 1 + 2 \lambda + 3 \mu + 6 \lambda^2 + 20 \lambda \mu + 15 \mu^2 \right) \frac{r_0^4}{r^4}\nn\\
 && +\left( \lambda + 3 \mu+ 6 \lambda^2 + 30 \lambda \mu + 30 \mu^2 \right)  \frac{r_0^8}{r^8} + \CO(\lambda^3, \mu^3, \frac{r_0^{12}}{r^{12}})   \nn\\
 &=& f_{\infty} \left[1-  \left( 1 + \lambda + 2 \mu + 3 \lambda^2 + 12 \lambda \mu + 10 \mu^2 \right)  \frac{r_0^4}{r^4} \right. \nn\\
 && \left. ~~~~~~~+  \left( \lambda + 3 \mu+ 5 \lambda^2 + 26 \lambda \mu + 27 \mu^2 \right)\frac{r_0^8}{r^8}  \right]  + \CO(\lambda^3, \mu^3, \frac{r_0^{12}}{r^{12}}) \ ,
 \ee  
where an overall $f_\infty=\lim_{r\to \infty} f(r)$ have 
been factored out since it just gets absorbed into the effective curvature $\tilde{L} = f_{\infty}^{-1/2} L$.     
The extra parameters $\mu$ and $\lambda$ allow us to independently 
vary the coefficients of powers of ${1\over r}$ in $f(r)$ in the large $r$ expansion. 
It is convenient to simply relabel these independent coefficients as\footnote{The bare  $f_0$ and the black hole mass depend on the terms $\mu$ and $\lambda$ in the Lagrangian, but these are unphysical as the Lagrangian can be changed by field redefinitions. The question is what happens when one parameterizes in terms of physical quantities, in our case they are OPE coefficients and operator dimensions. 
} 
\be
\label{fr}
f(r) = 1 - \frac{f_0}{r^4} - \frac{f_4}{r^8} - \frac{f_8}{r^{12}} - \dots \ .
\ee 
This matches precisely the forms \eqref{ourf}, \eqref{ourh} but with $f(r)=h(r)$.   

The parameter $f_0$ plays a special role in our analysis. It is fixed by the ratio between the dimension 
of the heavy operator $\Delta_H$ and the central charge $c$ in the dual $d=4$ CFTs:\footnote{Recall  \eqref{eq:f0def1} and $C_T= {40\over \pi^4} c$ \cite{Osborn:1993cr}.}
\be
f_0= {4 \pi^4 \over 3} {\Delta _H \over c} ~~~~~ (d=4)~\ .      
\ee 
The additional parameters $f_2, f_4,...$, represent other generally  coupling- and model-dependent quantities.

\subsubsection{Perturbative Solutions}

We now discuss stress-tensor contributions \eqref{STsol} 
in the background \eqref{bg1}, which corresponds to 
a generalization of quasi-topological gravity.
Here we again focus on the planar black hole.\footnote{See appendix \ref{app:spherical} for the spherical black hole case.} 

We find  
\be
&&\Phi^T = \Big({r\over  w^2}\Big)^\Delta \Big (1+G^T(r, w , \Hat \rho) \Big) \ ,  ~~~G^{T} ={1\over r^4} \sum_{i=0,4,8,...} {G^T_{i}(w, \y)\over r^i} \ ,
\ee 
where
\be
 G^T_{i}= \sum_{j=0,2,4,..}^{2+{i\over 2}} \a^{(j)}_{i}(w)  \y^j  
\ee
has the following polynomial forms: 
\be
G^T_{0}&=&\a^{(0)}_0(w)+ \a^{(2)}_0(w)\y^2\nn\\
&=& \sum_{i=-2}^{4} a_i w^i+\sum_{j=-2}^{2} b_j w^j  {\Hat \rho}^2\ , \\  
G^T_{4}&=&\a^{(0)}_2(w)+ \a^{(2)}_2(w)\y^2+ \a^{(4)}_2(w)\y^4 \nn\\
&=& \sum_{i=-4}^{8} a_i w^i+\sum_{j=-4}^{6} b_j w^j  {\Hat \rho}^2+\sum_{k=-4}^{4} c_k w^k  {\Hat \rho}^4\ , \\   
G^T_{8}&=& \a^{(0)}_8(w)+ \a^{(2)}_8(w)\y^2+ \a^{(4)}_8(w)\y^4+\a^{(6)}_8(w)\y^6  \ , \nn\\ 
&=&  \sum_{i=-6}^{12} a_i w^i+\sum_{j=-6}^{10} b_j w^j  {\Hat \rho}^2+\sum_{j=-6}^{8} c_k w^k  {\Hat \rho}^4+\sum_{l=-6}^6 d_l w^l  {\Hat \rho}^6 \ ,
\ee and so on.  
One can easily identify a general pattern. 
To keep expressions simple, we have not added subscript/superscript for the constant coefficients 
$a_i, b_j, c_k,..$ etc. 

At order ${\cal O} ({1\over r^4})$, we have
\be
\label{GT0}
G^T_{0}:
& &a_{-2}= - {\Delta \over 5} (f_0 + h_0)\ , \\  
&& a_{0}= -{3(h_0-f_0)-2  \Delta(f_0+h_0) \over 20}  \ ,\\  
&& a_{2}= - { f_0 (7 - 4 \Delta) + h_0 (\Delta-4)\over 60 (\Delta-1) }  \Delta \ , \\ 
&& a_{4}= - {f_0 (10 - 4 \Delta) + h_0 (\Delta-4) \over 120 (\Delta-2) } \Delta  \ ,  
\ee 
and 
\be
\label{GT0b}
&&b_{-2}= -{f_0 \over 5} \Delta  \ ,~~ b_{0}=- {f_0 \over 10} \Delta  \ , ~~ b_{2}=  -{f_0 \over 30}  \Delta  \ .  
\ee   The leading-order solutions in the planar black hole case are exactly the same as that in the spherical black hole case.  
It is necessary to impose the condition $h_0=f_0$ to remove poles at $\Delta=1,2$ and to match conformal block decomposition (see below).   
In this case, 
\be
G^T_{0}|_{h_0=f_0}:  
~~ a_{-2}= - {2 \over 5}  f_0  \Delta\ ,  
~~ a_{0}= {1 \over 5}  f_0  \Delta \ ,
~~ a_{2}=  {1 \over 20}  f_0  \Delta \ , 
~~ a_{4}= {1 \over 40} f_0  \Delta  \ .  
\ee  
The solutions \eqref{GT0b} remain untouched.

It is straightforward to obtain higher-order solutions using the computation scheme described above, but the 
expressions become increasingly cumbersome and we choose to only list the explicit solutions to the order ${\cal O}({1\over r^8})$ in appendix \ref{app:sol}.   
One may observe from the solutions listed in appendix \ref{app:sol} that the coefficients of 
the highest-power of $\y$ do not depend on $h(r)$, and these coefficients depend on $f(r)$ only through $f_0$. 
This figure will be directly related to the universal lowest-twist main result discussed in the next section with a more general setup.

\subsubsection{Conformal Block Decomposition}

Here we perform the $T^n$ conformal block decomposition to extract the corresponding OPE coefficients   
by taking a large $r$ limit on the bulk-to-boundary correlator. 

In $d=4$, the scalar 4-point function can be written as \cite{Dolan:2000ut, Dolan:2003hv} (See also \cite{Poland:2018epd} for a review.)
\be
\label{CB}
&&
\langle {\cal O}_H(0) {\cal O}_L(z,\bar z) {\cal O}_L(1) {\cal O}_H (\infty) \rangle
= \sum_{\Delta_T, J}  c_{\rm OPE} (\Delta_T, J) {B(z, \bar z, \tau, J)\over (z \bar z)^\Delta}  \ , 
\ee 
where $\tau\equiv\Delta_T-J$ and 
\be
\label{B}
&&B(z, \bar z, \tau, J)= {z \bar z \over z-\bar z}\nn\\
&&\times
\Big[z^{\tau + 2J\over 2} {\bar z}^{\tau-2 \over 2}
\, _2F_1\Big({\tau-2 \over 2},{\tau-2 \over 2}; \tau-2;  \bar z\Big)
\, _2F_1\Big({\tau + 2J \over 2},{\tau + 2J \over 2}; \tau + 2J;  z\Big) - (z \leftrightarrow \bar z) \Big]\ . \nn\\
\ee  
The relations between coordinates $t, \rho$ and $z,\bar z$ are\footnote{This relation is just the small $z,\bar{z}$ limit of \eqref{eq:zzbarTOttheta} with $\theta$ replaced by $\rho$.}  
\be
\label{transf} 
(t,\rho)=\frac{1}{2} \Big(-(z+\bar{z}), i (z-\bar z) \Big) \ .
\ee   
 In this planar black hole case, the following rescalings should be 
implemented to compute the OPE coefficients\footnote{See, for instance, \cite{Witten:2001ua}. 
}
\be
\label{rescaling}
(z,\bar z) \to ( {z\over r}, {\bar z\over r}) \ , 
~~ (f_i, h_i) \to ( r^{4+i} f_i,  r^{4+i} h_i), 
~~c_{\rm OPE} (\Delta_T, J) \to  r^{\Delta_T} c_{\rm OPE} (\Delta_T, J)  \ .
\ee 

First we consider the leading-order in OPE. 
From \eqref{GT0}--\eqref{GT0b}, 
\be
\label{G0general}
\lim_{r\to \infty} {G^T_0\over r^4}  
=- {z \bar z \Delta \over 120 (\Delta-2)} \Big( 2f_0 (z^2+3 z \bar z + \bar z^2)- f_0  \Delta (z+\bar z)^2+h_0 z \bar z (\Delta-4) \Big) \ . 
\ee  
On the other hand, \eqref{CB} and \eqref{B} with $\Delta_T=4$, $J=2$  give  
\be
\label{CB0}
c_{\rm OPE} (4, 2) z \bar z (z^2 + z \bar z + \bar z^2) \ .
\ee 
The consistency between \eqref{G0general} and \eqref{CB0} requires
$h_0=f_0$. We have
\be
\lim_{r\to \infty} {G^T_{0} \over r^4}  |_{h_0=f_0}=  {\Delta f_0\over 120} z \bar z (z^2 + z \bar z + \bar z^2) \ . 
\ee 
Thus, 
\be
c_{\rm OPE}(4,2)=  {\Delta \over 120} f_0 \ , 
\ee  
which is simply the same leading-order result obtained earlier with a spherical black hole. 
In what follows, we set $h_0=f_0$.  

Consider the next order with solutions listed in appendix \ref{app:sol}. 
By matching   
\be
&&\lim_{r\to \infty} {G^T_4\over r^8}  =c_{\rm OPE}(8,0)~  z^4 \bar z^4  
+ c_{\rm OPE}(8,2)~z^3 \bar z^3 \big(z^2 + z \bar z + \bar z^2\big)  \nn\\
&&~~~~~~~~~~~~~~~~~~ + c_{\rm OPE}(8,4)~z^2 \bar z^2 \big(z^4 + z^3 \bar z + z^2 \bar z^2 + z \bar z^3 + \bar z^4\big) \ ,
\ee  
we find 
\be
\label{80}
&&c_{\rm OPE}(8,0)={\Delta\over{201600 (\Delta -4)(\Delta -3) (\Delta -2)}}\nn\\
&&~~~~~~~~~~~~~~~~~~~ \times \Big[\big(7 \Delta ^4-45 \Delta ^3+100 \Delta ^2-80 \Delta +48\big) f_0^2 \nn\\
&&~~~~~~~~~~~~~~~~~~~~~~~~~~~~  +40 \left(\Delta ^3-3 \Delta ^2+20 \Delta +24\right) (2 f_4-h_4) \Big] \ , \\ 
\label{82}
&&c_{\rm OPE}(8,2)={ \Delta\over 201600 (\Delta -3) (\Delta -2)}\nn\\
&&~~~~~~~~~~~~~~~~~~~ \times \Big[ \big(7 \Delta ^3-23 \Delta ^2+22 \Delta +12\big) f_0^2+ 80 \left(\Delta ^2+3 \Delta +2\right) f_4\Big] \ , 
\ee
and the lowest-twist, 
\be
\label{84}
c_{\rm OPE}(8,4)=f_0^2  \frac{ \Delta (7 \Delta ^2+6 \Delta +4) }{201600 (\Delta -2)} \ .  
\ee  
While the condition $h_0=f_0$ must be imposed, we do not find a condition such as $h_4=f_4$ to be necessary.  

In the present high-temperature limit,  $c_{\rm OPE}(10+4k, J)=0$ with $k=0,1,2,3,4,...$. (Recall remarks below \eqref{eq:LTOps}.)
For reference, we list the next order's lowest-twist coefficient:
\be
c_{\rm OPE}(12,6)
=  \Delta  f_0^3 \frac{  1001 \Delta^4 +3575 \Delta^3 +7310\Delta^2 +7500 \Delta +3024}{10378368000 (\Delta -3) (\Delta -2)} \ . 
\ee 
As mentioned, the poles at integer $\Delta$ indicate mixing with double-trace modes. 

Observe that the above lowest-twist coefficients depend on $f_0$ (and $\Delta$) only. 
The higher-twist coefficients, \eqref{80}, \eqref{82}, are however explicitly contaminated by additional parameters such as $f_4, h_4$.  
We have explicitly computed the OPE coefficients at higher orders and 
the pattern that the lowest-twist coefficients are generally protected persists.  
In the next section, we will provide an all-orders proof of the universal lowest-twist without referring to a specific gravity model.  

We will next consider the geodesic approximation, which provides a useful check on the results obtained above. 

\subsubsection{Geodesic Approximation
}

At large $\Delta$, the two-point function can be approximated by the geodesic length $\sigma_g$:
\be
\label{logoo}
\lim_{\Delta\to\infty}\log \langle {\cal O}{\cal O}\rangle= -\Delta \sigma_g  + {\rm sub} {\rm leading~in~}\Delta \ .
\ee   Here we compute the geodesic length in a black hole background with higher-derivative corrections.
Start with\footnote{To consider a spherical black hole, one simply replaces $r^2 f$ and $r^2 h$ with $1+r^2 f$ and $1+r^2 h$, respectively. See \cite{Fidkowski:2003nf} for a similar geodesic computation in Einstein gravity.}  
\be
ds^2= {r^2 } f(r) dt^2+{dr^2 \over {r^2} h(r)} + r^2  \sum_{i=1}^3 d x^2_i  \ .
\ee
We remove $x_2$ and $x_3$ dependences using translational symmetry, and rename $x_1$ as $\rho$ in the following.
Writing $s=\int d s= \int {\cal L} d\tau$, one can first identify conserved quantities, momentum $l$ and energy $H$, 
for a geodesic: 
\be
l= \partial_{\dot \rho} {\cal L}={r^2\dot \rho }\ , ~~~~~ H= \partial_{\dot t} {\cal L}=r^2  f \dot t \ , 
\ee 
where $\dot \rho=\del_\tau \rho$, $ \dot t= \del_\tau t$. 
These two quantities are $h$-independent.  
One can next derive 
\be
\label{relr2}
\dot r^2 = h \big( r^2-{l^2} \big)-{h\over f} H^2 \ .
\ee  
To keep expressions simple,
instead of adopting the most general forms of $h$ and $f$, we set $f=h$ in this geodesic computation, and then \eqref{relr2} reduces to
$\dot r^2= f \big(r^2-{l^2}\big) -H^2$.

Denote $r^*$ as the turning point of the geodesic. 
The geodesic time and ``angle" are 
\be
\label{gt}
t (H,l)&=& 2 \int_{r^*}^\infty \frac{ H dr } {r^2 f \sqrt{f \big(r^2-l^2\big)-H^2}} \ , \\
\label{gp}
\rho(H,l)&=& 2 \int_{r^*}^\infty  \frac{ l dr}{r^2 \sqrt{f \big(r^2-l^2\big)-H^2}} \ .
\ee
The geodesic length is 
\be
\label{sigma}
\sigma_g (H, l)= 2 \int_{r^*}^\Lambda \frac{ dr}{\sqrt{f \big(r^2-l^2\big)-H^2}} \ ,
\ee 
where $\Lambda$ is an IR cut-off.  
The above expressions are functions of $H$ and $l$ and our task is to solve for the map 
\be
\Big(t (H,l), \rho (H, l)\Big)
 \to 
\Big(H(t,  \rho),  l(t,  \rho)\Big) 
\ee  
in a large $r$ expansion to obtain the geodesic length \eqref{sigma} as a function of $(t, \rho)$.  

Perturbatively, 
\be
f(r) = 1 - \epsilon  {f_0\over r^4} - \epsilon^2 {f_4\over r^8 }-\CO(\epsilon^3) \ , 
\ee 
where $\epsilon$ is added simply to keep track of the expansion order. 
Similarly,
\be
H=H_0 +\epsilon  H_1+ \epsilon^2 H_2+\CO(\epsilon^3)\ ,~ l=l_0 +\epsilon  l_1+ \epsilon^2  l_2+\CO(\epsilon^3)  \ ,~ r^*=r^*_0 +\epsilon  r^*_1+ \epsilon^2  r^*_2+\CO(\epsilon^3)  \ .
\ee 
First compute the turning point, which is determined by 
\be
f \left(r^2-l^2\right)-H^2|_{r=r^*}=0 \ .
\ee  
We find ($\zeta^2=H_0^2+l_0^2$)
\be
\label{zeta}
&& r^*_0=\zeta \ , \\
&& r^*_1= \frac{f_0 H_0^2}{2 \zeta^5}+\frac{H_0 H_1+ l_0 l_1}{\zeta}\ , \\ 
&& r^*_2= -\frac{9 f_0^2 H_0^4}{8 \zeta ^{11}} 
+\frac{\left(f_0^2+f_4\right) H_0^2}{2 \zeta ^9} 
-\frac{5 f_0 H_0^2 \left(H_0 H_1+l_0 l_1\right)}{2 \zeta ^7} 
+\frac{f_0 H_0 H_1}{\zeta ^5}\nn\\ 
&&~~~~~~ -\frac{H_0 \Big(2 H_1 l_0 l_1+H_0 \left(H^2_1-l^2_1\right)\Big)}{2 \zeta ^3}
+\frac{H_1^2+2H_0 H_2+2l_0 l_2}{2\zeta } \ ,
\ee 
and so on. The expressions become increasing complicated at higher orders and we shall not list them explicitly here. 

Next we solve for the map by plugging the above turning-point results into  \eqref{gt} and \eqref{gp}. 
We obtain  ($\eta^2= t^2+\rho^2$)
\be
\label{map}
&&({H_0\over t}, {l_0\over \rho}) = {2\over \eta^2} (1, 1)   \ , \\
&& ({H_1\over t}, {l_1\over \rho})= - \frac{ f_0 }{30} \Big( 3 t ^2+\rho^2 , t^2-\rho^2 \Big)  \ , \\  
&&({H_2\over t}, {l_2\over \rho})
=-\frac{\eta^2 }{12600}\Big(
20 f_4 \eta^2 (7 t^2+\rho^2) 
+f_0^2 (77 t^4-14 t^2 \rho^2-11 \rho^4) , \nn\\
&&~~~~~~~~~~~~~~~~~~~~~~~~~~  
20 f_4 \eta^2 (5 t^2-\rho^2) 
+f_0^2 (21 t^4-46 t^2 \rho^2+13 \rho^4)  \Big)   \ ,
\ee and so on. 
With the map, we finally compute the geodesic length.  
The result up to the order starting to contain $f_4$ is 
\be
\label{geodresult}
&&\sigma_g(t,\rho)
=2 \log \Lambda +2 \log  \eta
-\frac{\eta^2 }{120}  f_0 \left(3 t^2-\rho^2\right)\nn\\ 
&&~~~~~~ -{\eta^4 \over 100800} \Big[20 f_4\eta^2 \left(7 t^2-\rho^2\right) + f_0^2 \left(77 t^4-70 t^2 \rho^2+13 \rho^4\right)\Big]+ {\cal O} (f^3_0)\ . 
\ee

It is straightforward to consider higher-order corrections.  For instance,  the next-order correction is 
\be 
\label{nextordersigma}
&&\sigma_g(t,\rho)|_{{\cal O} (f^3_0)}=-{\eta ^6 \over 1297296000} \Big[9000 f_8 \eta ^4 \left(11 t^2-\rho ^2\right)+ 60 f_0 f_4 \eta ^2 \left(4587 t^4-3058 \rho ^2 t^2+ 347 \rho ^4\right)\nn\\
&&~~~~~~~~~~~~~~~~~~~~~~~~~~~~~~~~~~~~~ +f_0^3 \left(61677 t^6-94413 \rho ^2 t^4+40623 \rho ^4 t^2-5911 \rho ^6\right)\Big]\ .
\ee
As a consistency check, we have verified that the regularized geodesic length matches the leading large $\Delta$ limit of the 
boundary correlator.\footnote{Matching \eqref{geodresult} requires $G_4$ solution  in appendix \ref{app:sol}. 
In this geodesic computation, we set $f=h$ to have simpler expressions. 
Matching the next-order result, \eqref{nextordersigma}, requires solution $G_8$, which is rather long and we will not list it here.} 

Since the leading large momentum limit should correspond to the lowest-twist limit, one 
should be able to see also from the geodesic approximation that the result at large momentum 
depends only on  $f_0$.
Indeed, a large momentum implies a large turning point, as indicated by \eqref{zeta}, and at large momentum \eqref{sigma} becomes 
\be
\lim_{l \gg H} \sigma_g \sim \int_{r^*}^\Lambda \frac{ dr}{\sqrt{f(r) \big(r^2-l^2\big)}}  \ .
\ee  
We see the factor $f_0$ is selected out at large turning point.

Observe that, from the geodesic length expressions, \eqref{geodresult}, \eqref{nextordersigma}, 
only the factor $f_0$ is relevant at large $\rho$, with $t^2+\rho^2$ fixed.
This motivated us to look into a new limit: 
\be
\label{largey}
\y \to \infty    \ ,
\ee  
with $\hat t^2 + \y^2$ fixed. 
In the next section we will see that \eqref{largey} turns out to be a powerful limit in the bulk-to-boundary propagator.

\renewcommand{\theequation}{4.\arabic{equation}}
 \setcounter{equation}{0}
\section{Universal Lowest-Twist}
\label{sec4}

Up to this point, we have attempted to slowly gain intuition about the behavior of the holographic correlator 
by explicit computations in some specific limits (e.g. large $r$ with $w, \hat{\rho}$ fixed, geodesic approaximation, $d=4$ example).  
Now, we are ready to take on a more general case,\footnote{Here we still restrict 
to planar black holes and neglect additional matter fields in the bulk, and also focus on stress-tensor contributions with integer $\Delta$.
} 
and show that the leading-twist products of stress tensors are all universally determined by the same data as a single stress tensor, $\Delta \equiv \Delta_L$ and $f_0\sim {\Delta_H\over C_T}$. 

\subsection{Reduced Field Equation in General Dimensions  
}

We begin with the scalar field equation (\ref{eq:BulkEOMGen}) in a planar black hole metric.  
As before, we factor out the pure AdS propagator $(r/w^2)^\Delta$ and change variables to $r,w, \hat{\rho}$, in which case the 
equation of motion can be written as in (\ref{PDEwy}).   

Next, we need to identify a limit that isolates the lowest-twist contributions.  
To figure out what this limit should be, it helps to look back at the expansion (\ref{STsol}). 
The terms $G_i^T$ are the coefficients of $\frac{1}{r^{d+i}}$, per (\ref{eq:Gexpansion2}), and operators with $n$ 
stress tensors 
enter into the term $G_{d(n-1)}$.   At fixed $r,w$,  the term $G_{d(n-1)}^T$ grows at large $\hat{\rho}$ like $\hat{\rho}^{2n}$, reflecting 
the fact that with $n$ stress tensors one can make a primary operator with spin at most $2n$.  
In order to pick out the largest-spin term in each $G_i^T$, we shall take a limit where $\hat{\rho}$ becomes 
large with $   
\frac{G_{d(n-1)}^T}{r^{nd}} \sim \frac{\rho^{2n}}{r^{n(d-2)}}$ fixed.  
That is, we shall fix both $w$ and  
\be
\frac{\hat{\rho}}{r^{\frac{d}{2}}} =  \frac{\rho}{r^{\frac{d-2}{2}}}  \equiv  u \ ,
\ee
and then take $r\rightarrow \infty$.  
Note that for $d>2$, this scaling means that we are taking 
$\rho \rightarrow \infty$ at the same time as $r\rightarrow \infty$, so one might worry that we are losing any 
connection with the OPE limit.  However, one should think of this limit as first doing a series expansion of $G^T$ 
in powers of small $\rho$ with $r$ and $w$ fixed, followed by taking $r$ to infinity with $w$ and $u$ fixed.   
In fact, we will perform such an expansion explicitly when we proceed to solve for the lowest-twist component of $G^T$. 
In Euclidean space, keeping $w^2 = 1+r^2(\rho^2+t^2)$ fixed as $r$ and $\rho$ become large is somewhat formal, but 
in Lorentzian signature it is physical and corresponds to a lightcone limit, $\rho^2 -t^2 \sim 0$.  

Having identified the limit that extracts the lowest-twist stress tensors,  
we next derive the corresponding {\it reduced field equation}.     
Substituting into the equation of motion \eqref{PDEwy} a propagator of the form
\be
\Phi(r,w,u) \equiv \left( \frac{r}{w^2} \right)^\Delta  \Big( Q(w,u) +{\cal O} \big({1\over r}\big) \Big)\ ,   
\label{eq:LTlimitform}
\ee
and take the large $r$ limit, we find the resulting reduced field equation for $Q$ in general dimensions is given by  
\be
\label{PROOF}
&& u^{-2} (1-w^2)^{1-\frac{d}{2}} \partial_w \Big( w^{1-2\Delta} (1-w^2)^{d\over 2} \partial_w Q \Big) +  u^{-1} k_{-}  \partial_w \Big( w^{-2\Delta} k_{+} \partial_u Q\Big)  \nn\\
&&~~ - \big(\frac{d}{2}-1\big)^2w^{1-2\Delta} u \partial_u \Big(u^{-1} \partial_u Q \Big)+ f_0 \partial_w \Big( w^{-1} \partial_w (w^{-2\Delta} Q)\Big) =0 \ , 
\ee 
where 
\be
k_{\pm}= \big((d-2)w^2-d\big)^{{1\over 2} \pm {\Delta\over 2}} \ .
\ee
The reduced field equation \eqref{PROOF} manifestly depends only on $f_0$ in the large $r$ expansion of $f(r)$ and $h(r)$. (Even we take $h_0\neq f_0$, the reduced field equation still is $h_0$-independent.)

The equation \eqref{PROOF} allows one to simply solving for the coefficients of the highest-power in $\y$ 
part of the scalar solution (at a given $1/r$ order) and thus provides a consistent truncation on the bulk field equation.\footnote{ 
For instance, solving the reduced field equation in $d=4$ at the order ${\cal O}({1\over r^8})\sim u^4$ leads to 
``lowest-twist" solutions \eqref{c-4}-\eqref{c4} without needing to solve for non-universal pieces \eqref{a-4}-\eqref{a8} and \eqref{b-4}-\eqref{b6}.}

\subsubsection*{{\it Remarks:}}

When we took the form (\ref{eq:LTlimitform}), 
we did not have to allow for terms in $Q$ that scaled like positive powers of $r$ with $u,w$ fixed, at large $r$. 
This is nontrivial and thus requires an explanation. 
Recall first that the spatial coordinate $\rho$ scales like $u r^{\frac{d-2}{2}}$ in this limit.
Consider an individual term as a function of $w,\rho,$ and $r$:
\be
\frac{w^a \rho^\ell}{r^{a+\delta}} \ .
\ee
In order for the boundary-boundary two-point function 
to be finite, there must be at least as many powers of $r$ downstairs as upstairs when $t,\rho$ are held fixed, so $\delta \ge 0$. 
Therefore, at large $r$ with $u$ fixed, 
 the most rapidly growing terms are of the form
\be
\frac{w^a u^\ell r^{\ell \frac{d-2}{2}}}{r^a} \ .
\label{eq:wuterm}
\ee
Moreover, the dimension of the boundary operator corresponding to such a term is $\Delta_\CO = a+\ell$.  
In the planar limit, the only  $T^n$ primary operators that contribute are products of $T$s without 
derivatives, so their spin $\ell$ is at most twice their dimension divided by $d$, i.e. $\ell \le \frac{2}{d}\Delta_{T^n}  = \frac{2}{d} (a+\ell)$. 
Thus, the power of $r$ in the denominator of (\ref{eq:wuterm}) is at least as great as in the numerator. 

Let us emphasize again that there are also double-trace operators in addition to $T^n$ operators.  
Because the dimension of the double-traces is 
controlled by $\Delta$ (plus integers) whereas for the $T^n$s the dimension is always an integer, the powers of $r$ at fixed 
$w$ and $u$ for these two types of operators generically do not differ by integers.  Therefore, the equation of motion does not mix these 
two series, since the equation of motion involves derivatives and integer powers of $r$.    
An important loophole is the case where $\Delta$ for the light operator is itself a positive integer.   
In this case, the field equation does mix double-trace contributions and multi-stress-tensor contributions.  
More conceptually, for positive integer $\Delta$, some 
double-trace operators and $T^n$ operators are exactly degenerate, and one cannot separate out the two kinds of 
operators just by looking at their dimension.  This ambiguity manifests itself in various poles in the stress tensor OPE coefficients at 
positive integer $\Delta$; in this case, the contribution from the $T^n$s alone is 
singular, and should become regular after including the double-traces.

\subsection{Recursion Relation}

To solve the equation \eqref{PROOF}, we expand $Q$ in a series of powers of $u$:\footnote{This follows from the $\y$-expansion discussed in section \ref{nbs}.}     
\be
Q(w,u) = 1+u^2 Q_2(w) + u^4 Q_4(w) + \dots.
\ee
The equation of motion for $Q_2$ simplifies if we define
\be
Q_2(w) \equiv 
w^{2 d} \left(w^2-1\right)^{-\frac{d}{2}-1} \int^w dw' U_2(w') \ ,
\ee
in which case
\be
0 = w U_2'(w)
+ \left(\frac{d \left(w^2-2\right)}{w^2-1}-2 \Delta +1\right) U_2(w)  
-4 \Delta  (\Delta +1) f_0 \left(w^2-1\right)^{d\over 2} w^{-2 d-3}   
\ .
   \ee
 The solutions to this equation are 
 \be
 U_2(w) &=& w^{-2 d-3} \left(w^2-1\right)^{d\over 2} \left(c_1 w^{2 \Delta +2}
- 
2 \Delta  f_0\right) \ , 
 \ee
 where $c_1$ is an integration constant; as in section \ref{sec:LeadingD4Analysis}, $c_1$ should be set to 
zero in order to satisfy the standard boundary condition on the bulk-to-boundary propagator.  
Moreover, to avoid a singularity in $Q_2$ at $w=1$, we shall integrate $U_2$ from $w'=1$ to $w$:
 \be
 Q_2(w) &=&  - 2 \Delta  f_0 w^{2 d} \left(w^2-1\right)^{-\frac{d}{2}-1} \int_1^w dw' w'^{-2 d-3} \left(w'^2-1\right)^{d\over 2} \nn\\ 
  &=&- \frac{2  \Delta  f_0 }{d+2} w^{d-2} \left(w^2-1\right)^{-\frac{d}{2}-1}  \nn\\ 
&&\times \left( 
{2  w^{d+2}\Gamma^2 \left({d\over 2}+2\right) \over \Gamma (d+3)}
-\, _2F_1\left(-\frac{d}{2},\frac{d+2}{2};\frac{d+4}{2};\frac{1}{w^2}\right)\right) \ . 
 \ee
In even $d$, the above function  
has only a finite number of powers of $w$, but in odd $d$ there are an infinite number of terms in a $1/w$ series expansion.  

To read off the OPE coefficient from a single stress-tensor exchange, we re-expand the above $Q_2$ solution at large $r$ 
with instead $t,\rho$ fixed, and decompose the result into conformal blocks.  
Since $Q$, by definition, keeps track only of the lowest-twist contributions, we can keep only the leading power of $\bar{z}$ in the stress tensor conformal block. 
The resulting decomposition produces  ($\Delta_T=d$, $J=2$)  
\be
c_{\rm{OPE}}(d,2) = 
{f_0 \Delta\over d+2 }\frac{ \Gamma\left(2+\frac{d}{2} \right)^2}{ \Gamma\left(3 + d\right)}  \ .  
\label{eq:Pd2OPE}
\ee

The higher-order coefficients $Q_4(w), Q_6(w), \dots$ quickly become rather complicated functions of $w$.  
However, in even integer dimensions, we find that they all have simple finite series expansions in powers of $w$; for clarity, we will therefore restrict to even $d$.

In even integer dimensions, we identify the following series expansion structure of lowest-twist solution $Q$ in terms of $u$ and $w$:
\be
Q(w,u) = \sum_{n=0}^\infty \sum_m a_{n,m} u^{2n} w^{2m} \ , ~~~ \qquad d \in 2 \mathbb{N} \ , 
\label{eq:QwuSeries}
\ee where the range of $m$ is controlled by $n$ (see below), and $a_{n,m}$ are coefficients that depend on $f_0$ and $\Delta$ only.  
Substituting this expansion into the equation of motion (\ref{PROOF}) for $Q$ and matching coefficients, we obtain the following recursion relation: 
\be
&&a_{n,m} = \frac{1}{ 4(m-n d)} \Big[ \frac{\big(2(m+n-1)-n d\big)\big(2(m+n-1-\Delta)+(1-n)d\big)}{m-\Delta } a_{n,m-1} \nn\\
&&~~~~~~~~~~~~~~~~~~~~~~~~~~ - 4 f_0 \big(1-\Delta +m\big) a_{n-1,m+1} \Big]\ , \qquad d \in 2 \mathbb{N} \ . 
\label{eq:recursionrelation}
\ee
At $\CO(u^0)$, the solution is just $Q_0(w)=1$, so the coefficients satisfy an initial  
condition
\be
a_{0,m} = \delta_{m,0} \ .
\label{eq:RecursionBoundaryCondition1}
\ee
It is also straightforward to show by induction that if the series has a lowest power $w^{2m}$ at each value of $n$, it must be 
$m_{\rm{min}}=-n$.\footnote{This fact is obvious for $n=0$, since $Q_0(w)=1$.  Taking $Q_{2n}(w) \sim w^{\gamma_n}$ at small $w$, the    
 field equation  requires $\gamma_n = \gamma_{n-1}-2$ for the contributions from subsequent $Q_{2n}(w)$s to cancel against each other.} 
Moreover, in order to have a finite boundary value limit, $r\rightarrow \infty$ with $t,\rho$ fixed, the maximum power $w^{2m}$ at each $n$ must be $m_{\rm{max}}=\frac{d-2}{2} n$.    
These conditions, together with (\ref{eq:RecursionBoundaryCondition1}), are sufficient to fix the solutions to the recursion relation for $a_{n,m}$. 
 We have, for instance,
\be
(d=4) &:& a_{2,2} = f_0^2 \frac{\Delta  \left(7 \Delta ^2+6 \Delta +4\right)}{12600 (\Delta -2)} \ . \\
(d=6) &:& a_{2,4} = f_0^2\frac{\Delta  \left(429 \Delta ^3+677 \Delta ^2+1394 \Delta +600\right)}{16816800 (\Delta -4) (\Delta -3)}  \ .
\ee 
One can use the recursion relation \eqref{eq:recursionrelation} to compute the universal lowest-twist coefficients to higher orders.\footnote{In our convention, 
$a_{n,n} =(-4)^{J\over 2} c_{\rm{OPE}} (\tau_{\rm{min}}, J)$ in $d=4$.   
}

\subsection{Lowest-Twist ``Thermalization"}
\label{sec:LTTherm}

The contribution to the boundary two-point function is given by the  $m=\frac{d-2}{2} n$ terms. We have (still in even $d$)
\be
(z \bar{z})^\Delta \<\CO_L \CO_L\> \supset  \sum_{n=0}^\infty a_{n,\frac{d-2}{2} n} \left( \rho^2 (t^2 + \rho^2)^{\frac{d-2}{2}} \right)^n  \ ,
\ee
where the notation $\supset$ denotes that this expression simply focuses on capturing the lowest-twist contributions.  
We can next eliminate $\rho$ in favor of $\bar{z},t$ using \eqref{transf}, and in the small $\bar{z}$ limit the relation is simply
\be
t^2 + \rho^2 \sim - 2 t \bar{z} \ , ~~ \qquad \rho^2 \sim - t^2  \ .
\ee
We then have the leading small $\bar{z}$ contribution 
in $t, \bar{z}$ coordinates:
\be
(z \bar{z})^\Delta \<\CO_L \CO_L\> \supset \CV_d(t,\sigma) 
&\equiv& \sum_{n=0}^\infty a_{n,\frac{d-2}{2} n} \left( -t^d \left(-\frac{2\bar{z}}{t}\right)^{\frac{d-2}{2}} \right)^n \nn\\
&=& \sum_{n=0}^\infty a_{n,\frac{d-2}{2} n} \left( -t^d \sigma^{\frac{d-2}{2}} \right)^n  \ , ~~~~\sigma \equiv -\frac{2\bar{z}}{t}  \ .
\ee
Aside from making the expressions more compact, the variable $\sigma$ defined above is convenient since it 
parameterizes the angle away from the lightcone of the path made when one varies $t$ with $\sigma$ fixed.  
By contrast, fixed $\bar{z}$ with varying $t$ sweeps out paths that are parallel to the lightcone.  

In order for the lowest-twist operators to dominate the 
two-point function,  
we need $\bar{z} \ll z$, which means 
\be
t - i \rho \sim 0 \ , ~~~\sigma \approx 1- \frac{i \rho}{t} \sim 0 \ .
\ee 
 Naively, this limit 
looks trivial, since only the $n=0$ term in the sum is nonzero at vanishing $\sigma$.  
However, the coefficients $a_{n, \frac{d-2}{2}n}$ are proportional to $f_0^n$, so by inspection we shall keep the full functional dependence in the limit
\be
 \sigma \rightarrow 0, \quad f_0 \rightarrow \infty, \quad f_0  \sigma^{\frac{d-2}{2}} t^d~ \textrm{ fixed} \ .
\ee
Since $f_0$ is being taken large, 
 this is a large temperature limit. 

As a check, consider the case $d=2$.  It is straightforward to check by explicit comparison that the coefficients $a_{n,0}$ reproduce the following function:
\be
\CV_2(t) = \sum_{n=0}^\infty a_{n,0} (-t^2)^n \stackrel{d=2}{=} \left( \frac{ \sqrt{f_0} ~t}{\sin \big( \sqrt{f_0} ~t\big)} \right)^\Delta \ , 
\label{eq:2dBlock}
\ee
which is indeed the correct result \cite{Fitzpatrick:2014vua,Fitzpatrick:2015zha}.\footnote{To help with the 
comparison, note that from (\ref{eq:Pd2OPE}) and the fact that in $d=2$ we have 
$c_{\rm{OPE}}(d,2) = \frac{\Delta_L \Delta_H}{C_T}$,  
we have $f_0 =  \frac{24 \Delta_H}{C_T} =  (2 \pi T)^2$,     
whereas at small $\bar{z}$ we have $t \sim \frac{1}{2} \log(1-z)$.   
So, $\sqrt{f_0} t \rightarrow  \pi T \log(1-z)$.}    
The form \eqref{eq:2dBlock} matches the two-point function for the
light operator in a $d=2$ CFT at finite temperature \cite{Cardy_1984, CARDY1986186}. 

Another check is that $\log(\CV_d)$ should grow linearly in $\Delta$ at large $\Delta$.\footnote{Moreover, $\log(\CV_d)$ should be  
$-\Delta$ times the geodesic length in the large angular momentum limit. } 
We have found this to be true of the coefficients $a_{n,\frac{d-2}{2}n}$ up to high order.   
For instance, in $d=4$,  
\be 
\lim_{\Delta\rightarrow \infty} \frac{1}{\Delta} \log(\CV_4) 
= - \frac{f_0 v}{30}  
+\frac{f_0^2 v^2}{630} 
- \frac{1583 f_0^3 v^3}{10135125}  
+\frac{3975313 f_0^4 v^4}{192972780000} + \dots ,
\ee  
where $v\equiv -t^4 \sigma$.

We have not found a general closed form expression for $a_{n,\frac{d-2}{2}n}$ for general $n$ and $d$.  
However, there are a few special cases where we can find closed-form expressions, which shed light on the general case.  

\begin{figure}[t!]
\begin{center}
\includegraphics[width=0.6\textwidth]{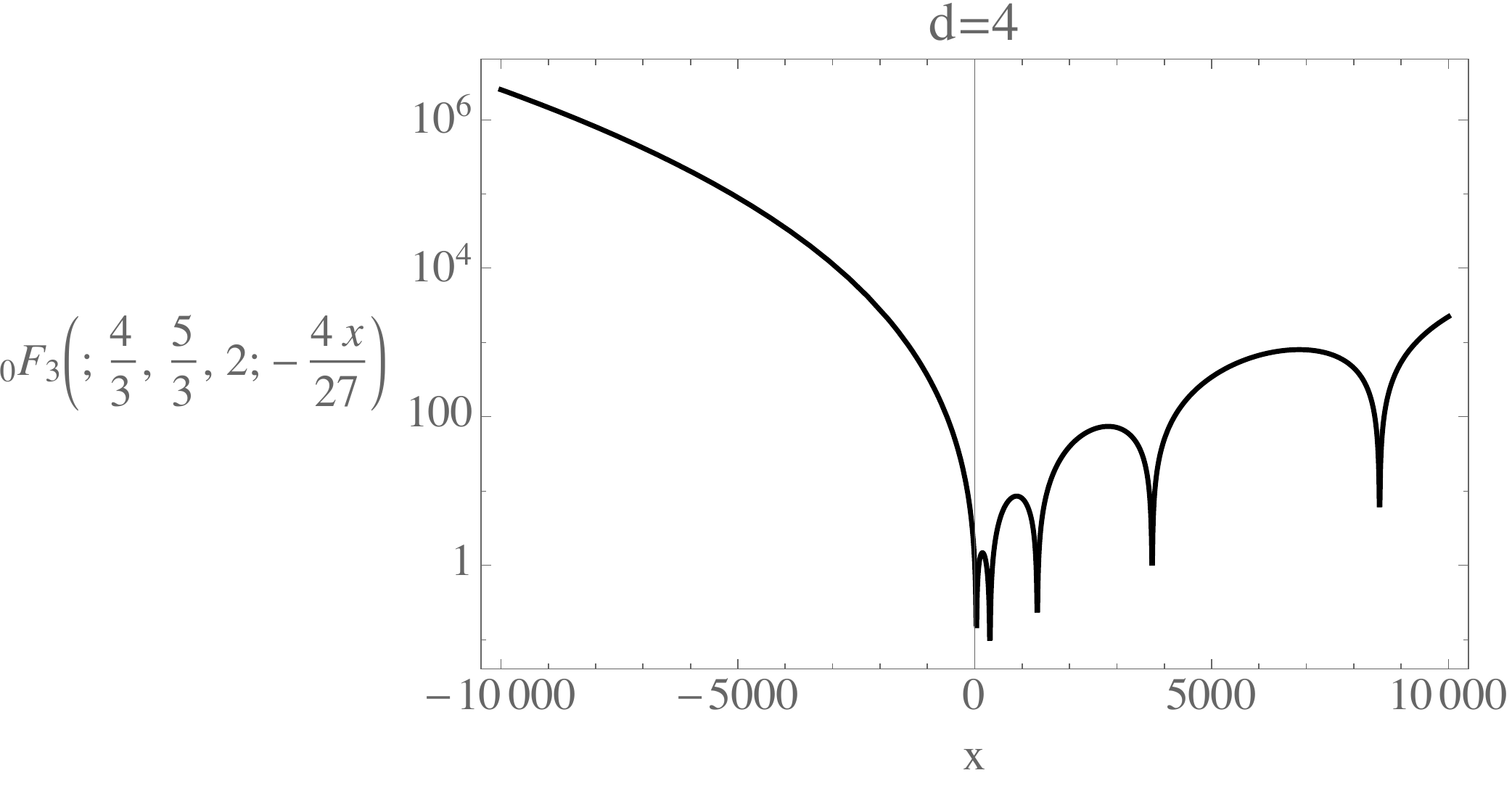}
\caption{$\CV_d$ in $d=4$ and $\Delta=-1$; eq(\ref{eq:CV4DeltaMinusOne}). }
\label{fig:Hyper}
\end{center}
\end{figure}

One such limit is partly motivated by the fact that in $d=2$, at $\Delta=-1$, $\CV_2$ is essentially just a sinh function, which has relatively simple series coefficients.  
Interestingly, at $\Delta=-1$, the $a_{n, \frac{d-2}{2}n}$ coefficients simplify in higher dimensions as well and take the following form:\footnote{More generally, simplifications occur at negative integer $\Delta$.  
A formula that appears to hold for even $d$ and $\Delta=-1, -2, \dots, -\frac{d+2}{2}$ is
\begin{equation}
a_{n,\frac{d-2}{2}n} =    \left( \prod_{j=1}^{\frac{d}{2}-1} \frac{1}{\left( \frac{2(j-\Delta)}{d-2}\right)_n\left( \frac{d-2}{2} \right)^n}\right) \left( \prod_{k=\frac{d}{2}}^{d} \frac{1}{\left( \frac{2(k-\Delta)}{d+2}\right)_n \left( \frac{d+2}{2} \right)^n}\right) 
\left( \frac{ f_0 \Gamma^2(\frac{d}{2}+1) \Gamma(1+d-\Delta)}{\Gamma(2+d)\Gamma(-\Delta)} \right)^{n} \ . 
\label{eq:DeltaGenCoefficients}
\end{equation}
}
\be
a_{n, \frac{d-2}{2}n} \stackrel{\Delta=-1}{=}   \left( \prod_{j=2}^{\frac{d}{2}} \frac{1}{\left( \frac{2j}{d-2}\right)_n\left( \frac{d-2}{2} \right)^n}\right) \left( \prod_{k=\frac{d}{2}+1}^{d+1} \frac{1}{\left( \frac{2k}{d+2}\right)_n \left( \frac{d+2}{2} \right)^n}\right) 
\left( f_0 \Gamma^2(\frac{d}{2}+1)\right)^{n} \ . 
\label{eq:DeltaMinusOneCoefficients}
\ee
These are, by definition, the series coefficients of hypergeometric functions ${}_0F_{d-1}$.  
For instance, with  $d=4$ and $\Delta=-1$,
\be
\label{CV4}
\CV_4(t,\sigma)|_{\Delta=-1}    
&=& \sum_{n=0}^\infty a_{n,n} (-t^4 \sigma)^n|_{\Delta=-1}  \nn\\
&=& \sum_{n=0}^\infty \frac{2^{2 n+1} \left(- f_0 t^4 \sigma \right){}^n}{\Gamma (n+2) \Gamma (3 n+3)}    
= {}_0F_3\left( \left\{\frac{4}{3} ,  \frac{5}{3}, 2\right\}, - \frac{ 4 f_0 t^4 \sigma}{27} \right) \ .  
\label{eq:CV4DeltaMinusOne}
\ee
We plot this function in Fig. \ref{fig:Hyper}.  At large $t$, it behaves like an exponential:
\be
{3^{\frac{1}{8}} \over 2^{\frac{9}{4}} \sqrt{\pi } \left( -f_0 t^4 \sigma\right)^{7/8}} ~ e^{\frac{4 \sqrt{2} (-f_0 t^4 \sigma)^{\frac{1}{4}}}{3^{3/4}}}\ . 
\ee
It is tempting to simplify $(t^4)^{\frac{1}{4}} \rightarrow t$ in the above expression so that $\CV_4(t,\sigma)$ 
at late times simply contains an exponential linear in $t$, $V_4 \sim e^{A t}$.  However, one must be more careful 
if one wants to analytically continue in $t$, say from Euclidean time (which we are using) to Lorentzian time $t_L = i t$.  
The original series was a convergent series in $t^4$ and therefore must be invariant under $t \rightarrow i t$.  
The issue is that there are multiple saddle points of the hypergeometric function, and subleading ones become 
leading under analytic continuation. This kind of feature is already present in $d=2$, where the asymptotic  
large $x$ behavior of $\sinh(x)$ is $\sim e^x$, but unlike $e^x$, $\sinh(x)$ does not decay at large negative $x$. 

To be more explicit about the correct asymptotic form, we can derive the large $t$ 
behavior of $\CV_4$ directly from its series expansion using \eqref{CV4}.  
At large $n$, still in $d=4$ with $\Delta=-1$,  
\be
a_{n,n}  \sim 
\frac{ f_0^n}{(4n)!}  \ .  
\ee
We have dropped an irrelevant overall prefactor 
and polynomial $n^{-7/2}$ dependence. 
The $(4n)!$ in the denominator 
indicates that these are the series coefficients of a sum of four exponentials:
\be
\sum_{n=0}^\infty  \frac{(t^4)^n}{(4n)!}   
= \frac{1}{4}\sum_{k=1}^4 e^{i^k t} \ .
\ee
Consequently, a more accurate asymptotic form of $\CV_4(t,\sigma)$ is     
$\sim \sum_{k=1}^4 \exp\left(  A_4 f_0^{\frac{1}{4}}  \sigma^{\frac{1}{4}}  i^k t \right)$.  

Similar analyses are straightforward for $\Delta=-1$ in any even dimension, starting from the coefficients (\ref{eq:DeltaMinusOneCoefficients}).  We find
\be
\CV_d(t,\sigma) \sim \sum_{k=1}^d \exp \left( A_d f_0^{\frac{1}{d}} \sigma^{\frac{d-2}{2d}} e^{\frac{2\pi i k}{d}} t\right)\ ,
\ee
where $A_d$ is a numeric factor. 

Do the above asymptotic forms imply certain thermalization  in higher dimensions? 
Observe that the temperature of an infinitely large AdS-Schwarzschild black hole is given by 
\be
\label{Tem}
\lim_{f_0\to \infty} T_{\rm AdS-Schw} = 
\frac{d}{4 \pi} 
(f_0)^{\frac{1}{d}} + {\cal O} (f^{-1}_0)\ , 
\ee
which is exactly the power of $f_0$ that has appeared in front of $t$ in our asymptotic expansion! 
However, the 
actual temperature of the black hole shall depend on the form of $f(r)$ near the horizon rather than just the $f_0$ term in 
the large $r$ expansion. It is therefore not clear in what sense, if any, $T_{\rm AdS-Schw}$ can be interpreted as a conventional temperature here.\footnote{
For instance, even in Gauss-Bonnet gravity where only a quadratic curvature term is added to the Lagrangian (with coefficient $\frac{\lambda}{2}$), the 
black hole temperature depends on both $f_0$ and $\lambda$ \cite{Myers:2010ru,Myers:2010jv}:  
\be
\lim_{f_0\to \infty} T_{\rm GB} = \frac{1}{ \pi} (f_0)^{\frac{1}{4}}  \left[ \frac{1-4\lambda + \sqrt{1-4 \lambda}}{2} \right]^{1\over 4}, \qquad (d=4) \ , 
\ee
where $f_0$ is still defined as the coefficient of the $1/r^4$ term in $f(r)$.
}

\subsection{Convergence Radius}

One way of saying why $\Delta=-1$ is convenient in $d=2$ is that in this case there are no poles or 
branch cuts in $\CV_2$, and the series expansion in $t$ has infinite radius of 
convergence.\footnote{Another distinction is that $\Delta=-1$, and in fact $\Delta=-n$, operators have null states 
in their descendants under the global conformal algebra.  In $d=2$, such operators have interesting shortening conditions under the  Virasoro 
algebra at infinite $c$ \cite{Chen:2016cms}.  The reason the Virasoro algebra appears even at infinite $c$ is that there is an $\CO(c)$ 
enhancement due to the dimension of the heavy state background.  It is tantalizing to suppose that a related mechanism may be at work in higher dimensions as well.}
More generally, $\CV_2$ has infinite radius of convergence for negative integer $\Delta$.  
Our series expansions for $\Delta=-1$ at $d>2$ also had infinite radius of convergence, but  we expect that, for generic $\Delta$, there should be a finite radius of convergence, as in the $d=2$ case. 

\begin{figure}[t!]
\begin{center}
\includegraphics[width=0.32\textwidth]{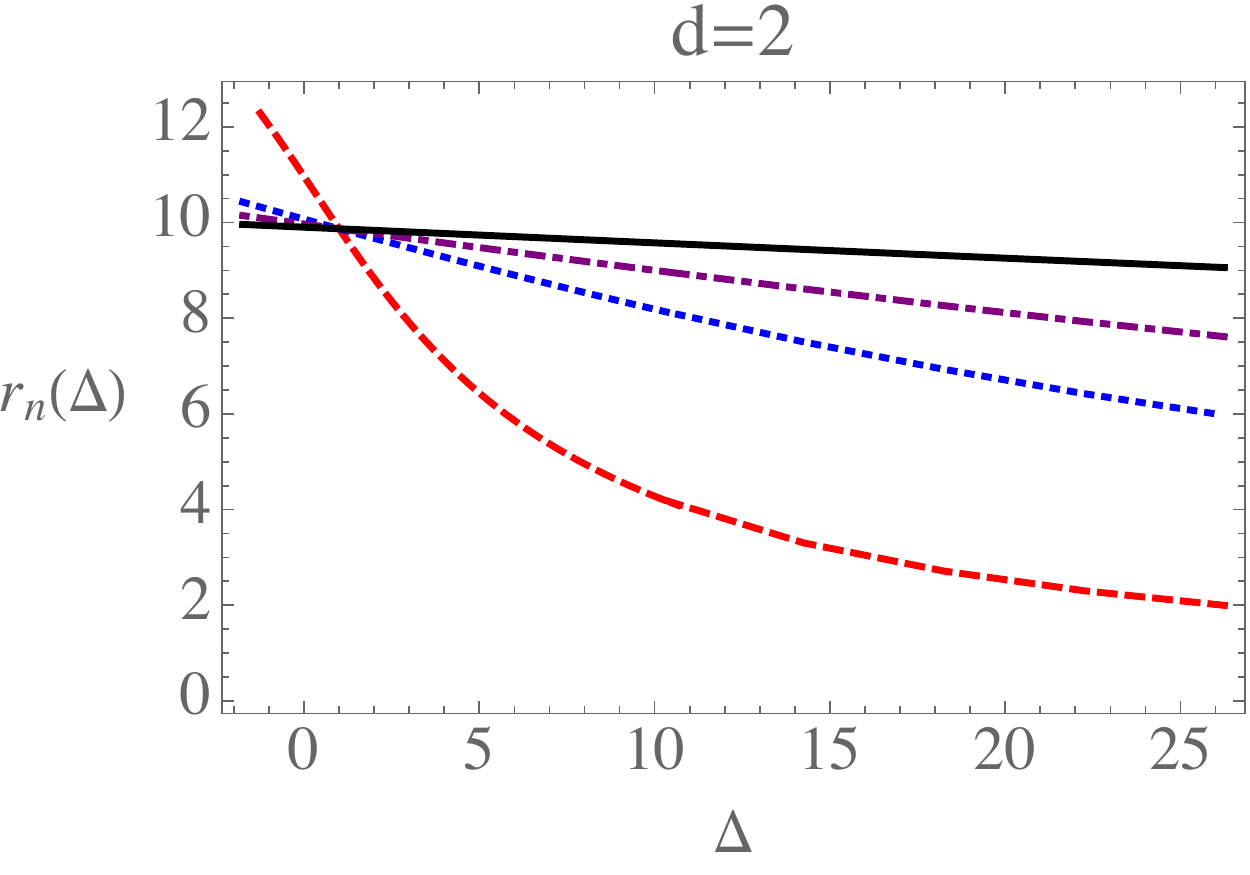}~~
\includegraphics[width=0.33\textwidth]{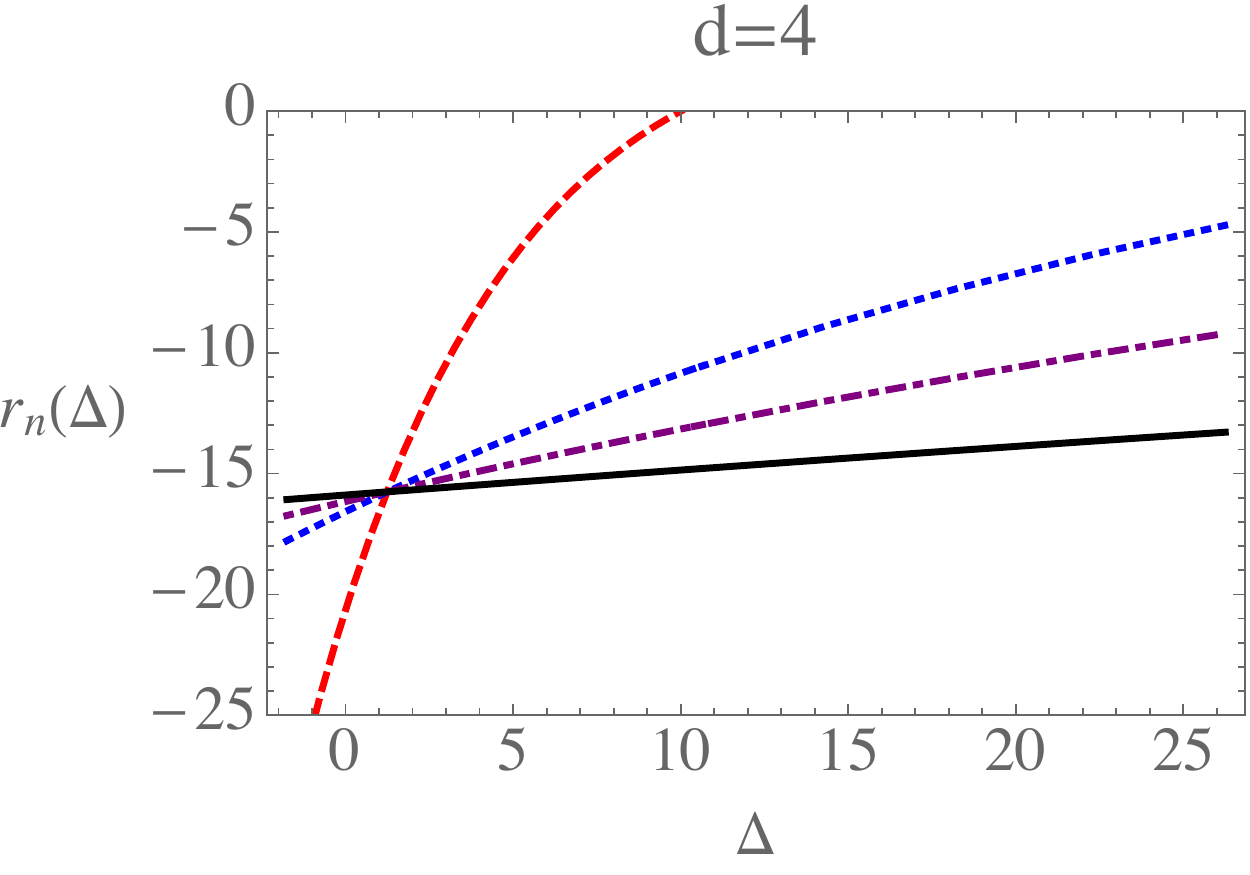}~~
\includegraphics[width=0.33\textwidth]{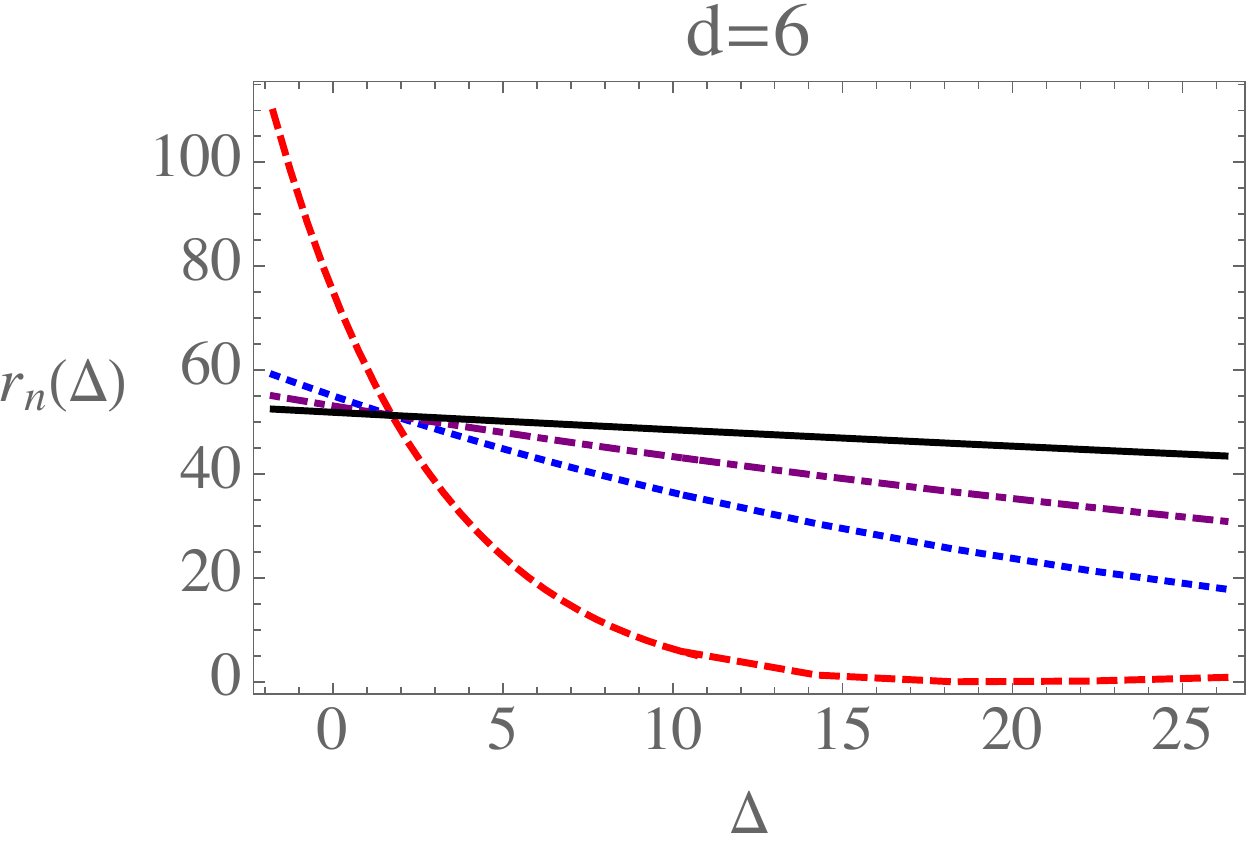}
\caption{Ratio $r_n(\Delta)$ for $d=2,4,6$ (left, middle, right)  at various $n$: $n=10$ (red, dashed), $n=50$ (blue, dotted), $n=100$ (purple, dot-dashed), and $n=300$ (black, solid). The curves appear to flatten as $n$ increases and approach a constant value $\lim_{n\rightarrow \infty} r_n(\Delta) = 9.87,  -15.756$, and $50.302$ at infinite $n$ in $d=2,4$ and $6$, respectively. 
The curves do not all intersect at a single point, though they appear to do so by eye; in the limit $n\rightarrow \infty$, $r_{n}(\Delta)$ and $r_{n-1}(\Delta)$  
intersect at $\Delta= 1,{5\over 4},{7\over 4}$ for $d=2,4,6$, respectively. 
}
\label{fig:ConvRate}
\end{center}
\end{figure}

We can investigate the radius of convergence  numerically by looking at the ratio of neighboring coefficients,
\be
r_n(\Delta) \equiv (- f_0) \frac{ a_{n-1,\frac{d-2}{2} (n-1)}}{a_{n, \frac{d-2}{2}n}} \ ,  
\ee
at large $n$. For $d=2,4$ and $6$, this ratio is plotted as a function of $\Delta$ at $n=10,50,100$ and $300$ in 
Fig. \ref{fig:ConvRate}.  As $n$ increases, the curves become increasing flat, especially near $\Delta=1,\frac{5}{4},$ 
and $\frac{7}{4}$ for $d=2,4$ and 6,  respectively, where the change in the curves as $n$ increases is minimal.   

\begin{figure}[t!]
\begin{center}
\includegraphics[width=0.32\textwidth]{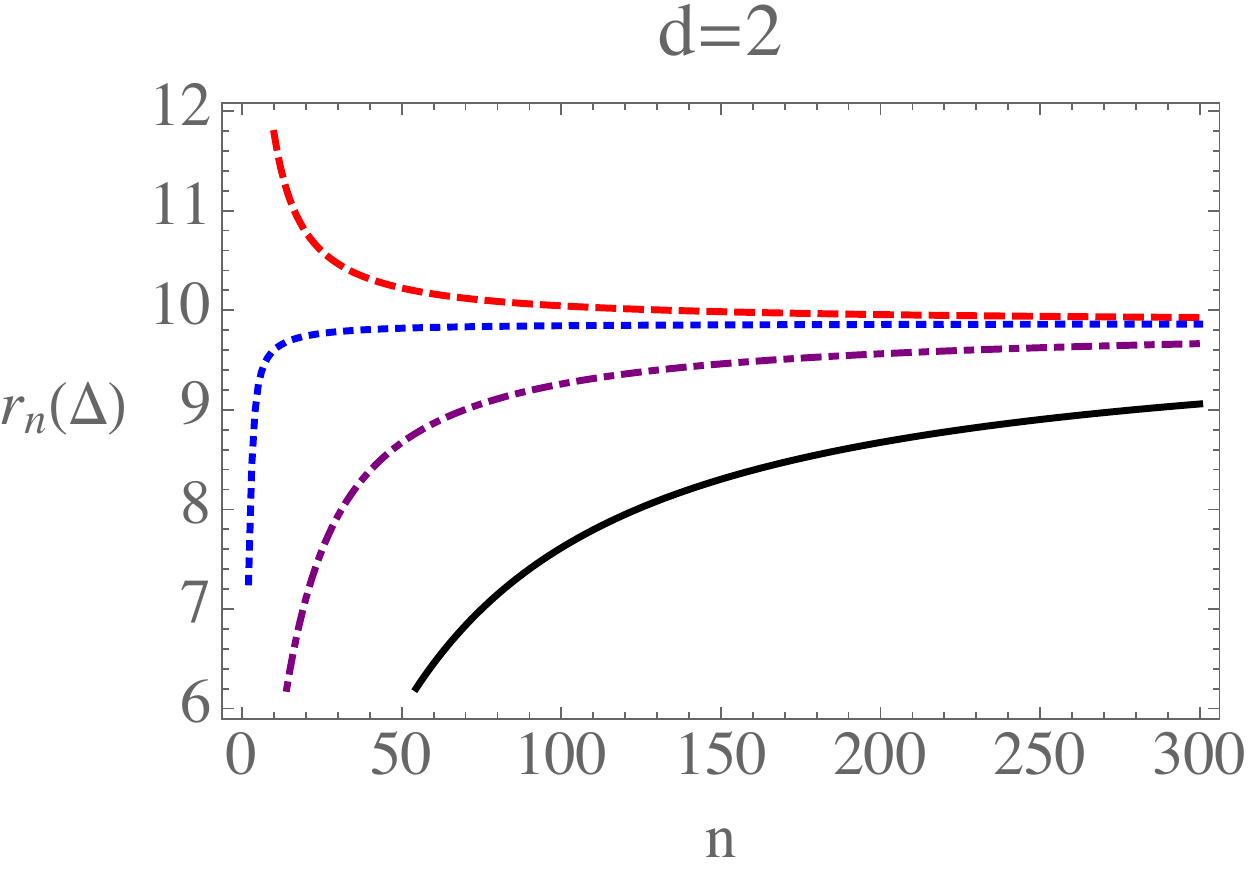}~~
\includegraphics[width=0.33\textwidth]{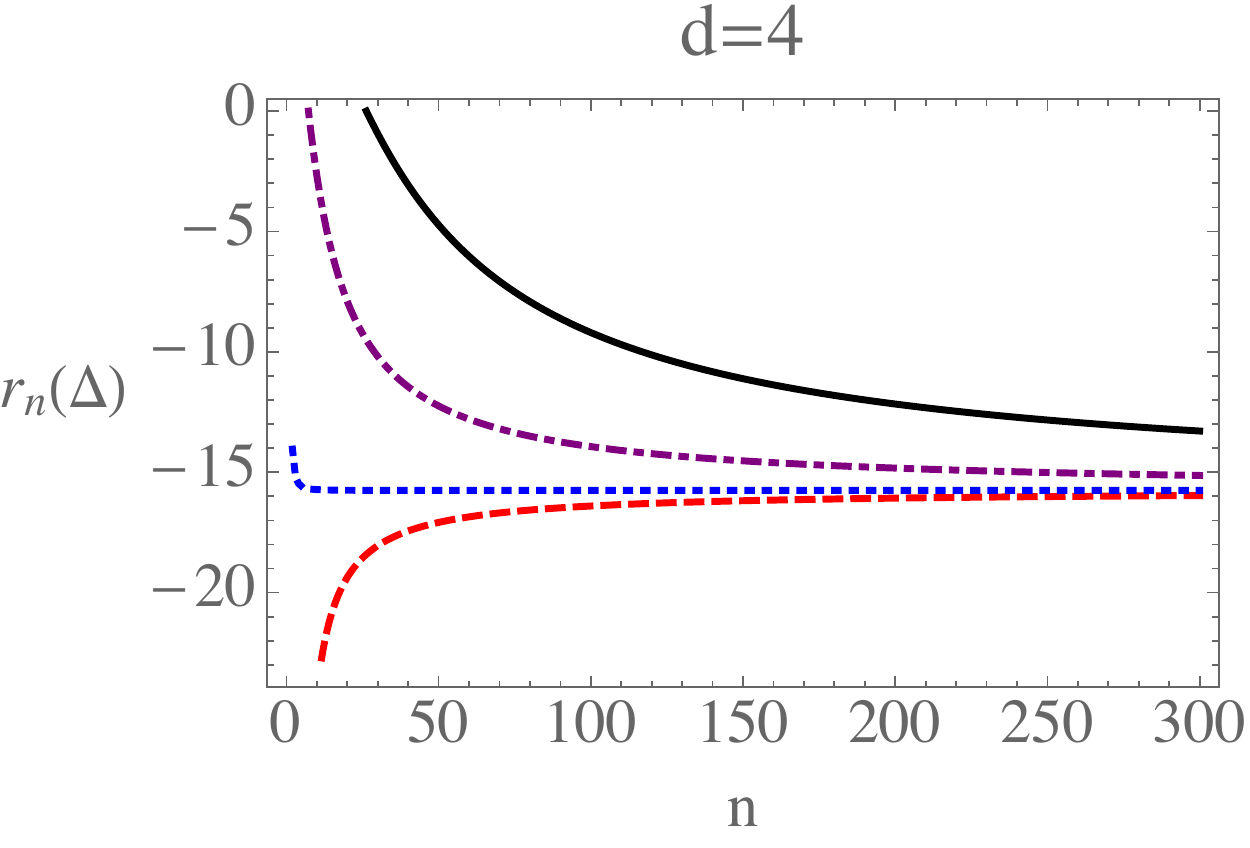}~~
\includegraphics[width=0.32\textwidth]{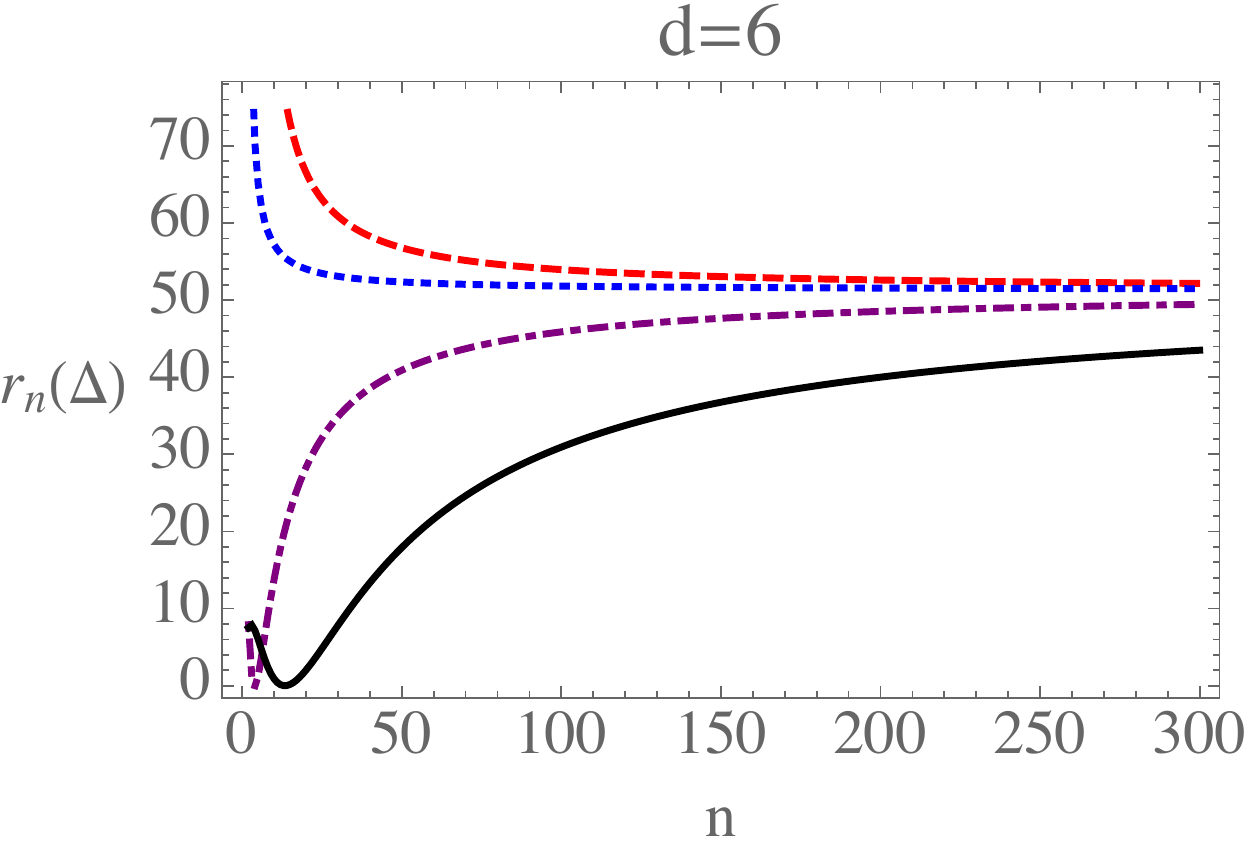}
\caption{Ratio $r_n(\Delta)$ for $d=2,4, 6$ (left, middle, right) at various $\Delta$: $\Delta=-0.75$ (red, dashed), $\Delta=1.25$ (blue, dotted), $\Delta=7.25$ (purple, dot-dashed), and $\Delta=26.25$ (black, solid). The curves appear to converge to the same value at infinite $n$.}
\label{fig:ConvRate2}
\end{center}
\end{figure}

In the case $d=2$, we know from the analytic result  (\ref{eq:2dBlock}) that  in the limit $n\rightarrow \infty$, the ratio 
approaches a constant value $\lim_{n \rightarrow \infty} r_n= \pi^2 \approx 9.87$, and we find that the numeric results suggest this 
behavior holds in higher $d$ as well, with $\lim_{n \rightarrow \infty} r_n \approx  -15.756$ for $d=4$  ($50.302$ for $d=6$).  
This constant ($\Delta$-independent) value $|r_\infty|$ should be the radius of convergence as a function of $f_0 t^4 \sigma$ for generic $\Delta$. 
This can be seen from Fig. \ref{fig:ConvRate} and Fig. \ref{fig:ConvRate2}.

\begin{figure}[t!]
\begin{center}
\includegraphics[width=0.5 \textwidth]{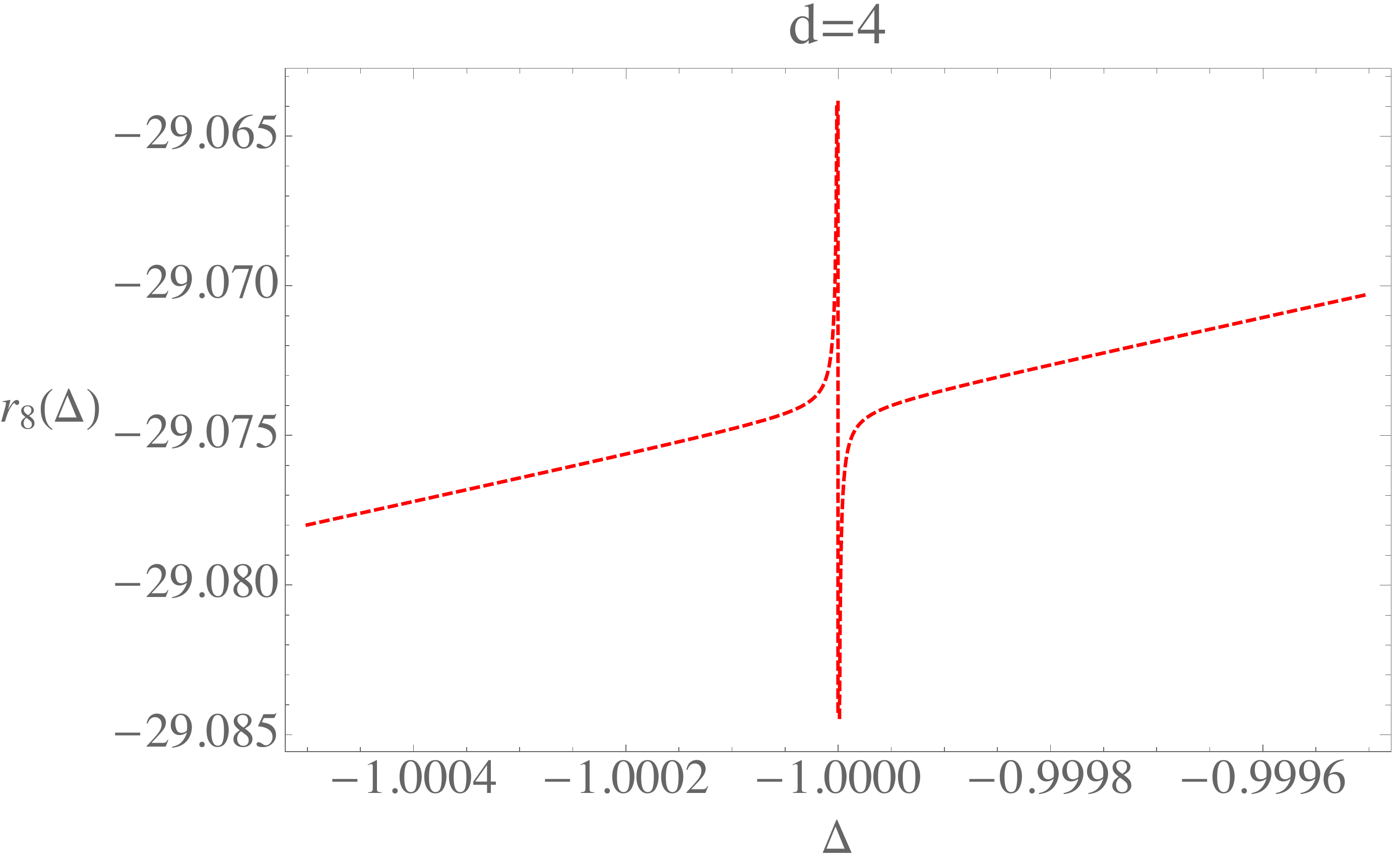}
\caption{Ratio $r_8(\Delta)$ for $d=4$ zoomed in near $\Delta=-1$. }
\label{fig:spike}
\end{center}
\end{figure}

In fact, the convergence of $r_n$ as a function 
of $n$ is rapid enough that the limit $r_\infty$ can be computed to a large number of digits and seen to match with high precision the following analytic form in even $d$: 
\be 
r_\infty =  \left(\frac{2-d}{2+d}\right)^{\frac{d-2}{2}} \left( \frac{2 B\left(\frac{1}{2},\frac{1}{d}\right)}{d} \right)^d \ ,
\label{eq:convergenceradius}
\ee
where $B$ is the beta function.\footnote{We have checked this form holds to better than one part in $10^{8}$ for $d=2, 4,6, \dots, 16$.} 

One may ask how such a flattening of the function $r_\infty(\Delta)$ is consistent with the infinite radius of convergence we saw at 
$\Delta=-1$ above.  The answer is that the ratio $r_n$ has a very sharp feature near $\Delta=-1$ that is invisible in the numeric plots 
in Fig. \ref{fig:ConvRate}. To better see this feature, we have zoomed in on $\Delta=-1$ at $n=8$ in Fig. \ref{fig:spike} (at higher $n$, the 
feature becomes so narrow that plotting it accurately becomes difficult).

\begin{figure}[t!]
\begin{center}
\includegraphics[width=0.7\textwidth]{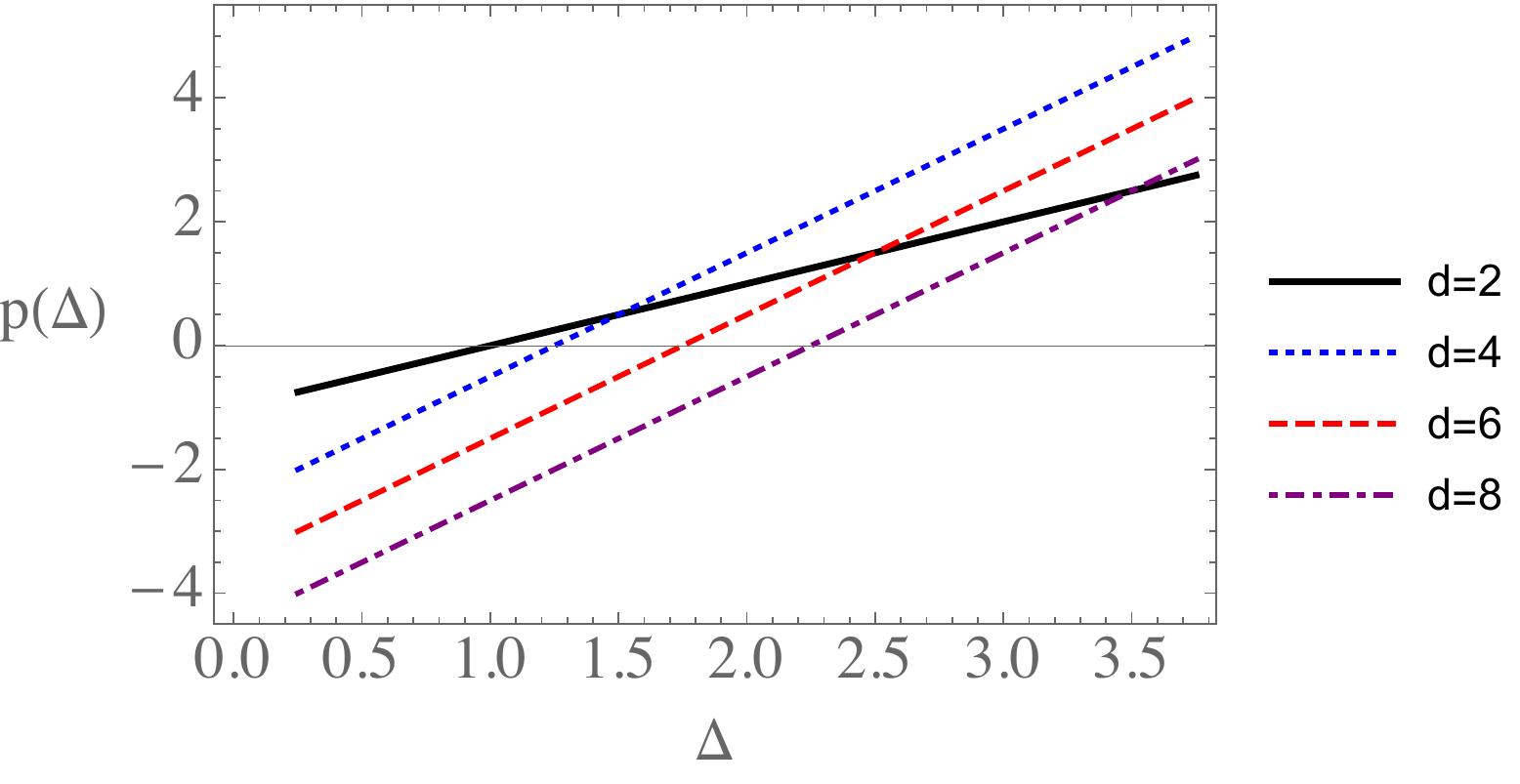}
\caption{Power $p(\Delta)$ for the large $n$ behavior of the coefficients $a_{n, \frac{d-2}{2}n} \approx A r_{\infty}^{-n} n^{p}$, for $d=2,4,6,8$.  For $d=2$, $p(\Delta) = \Delta-1$, whereas for $d=4,6,8$, $p(\Delta)= 2 \Delta- \frac{d+1}{2}$.  }
\label{fig:SingularExp}
\end{center}
\end{figure}

To learn more about the behavior of $\CV_d$ near its radius of convergence, we may fit the behavior of $a_{n, \frac{d-2}{2}n}$ at large $n$ to the form
\be
a_{n,\frac{d-2}{2}n} \approx A r_\infty^{-n} n^{p(\Delta)} \ ,  
\ee
where $A$ and $p$ are parameters determined by the fit.\footnote{We can improve the accuracy by allowing a few subleading terms as well, i.e. we take the form $r_\infty^{-n} n^p (A + \frac{B}{n} + \frac{C}{n^2} + \dots)$.}  By performing this fit at each value of $\Delta$, we obtain the exponent $p(\Delta)$ as a function of $\Delta$ shown in Fig. \ref{fig:SingularExp}.  For $d=2$, the exponent follow $p(\Delta)=\Delta-1$, which we also know analytically from the exact result (\ref{eq:2dBlock}). By contrast, for $d=4,6,8$  we numerically find the behavior is 
\be 
p(\Delta)= 2\Delta-\frac{d+1}{2} \ .
\ee 
Near the edge of the convergence radius, $\CV_d$ thus has  
 the form  
\be
\label{edgeform}
\sum_{n=0}^\infty a_{n, \frac{d-2}{2} n}\gamma^n \sim \left( 1 - \frac{\gamma}{r_\infty}\right)^{-p(\Delta)-1}  \ ,
\ee  
where $\gamma\equiv -t^d \sigma^{\frac{d-2}{2}}$.

\renewcommand{\theequation}{5.\arabic{equation}}
 \setcounter{equation}{0}
\section{Discussion and 
Future Directions}
\label{sec5}

The aim of this paper is to convey a message: the 
lowest-twist OPE coefficients of the $T^n$ conformal blocks in any dimensional CFTs (with or without supersymmetry) 
are universally protected, at least in the large central charge limit.\footnote{We assume no additional matter fields in this work and 
largely focus on the high-temperature limit.}  

We have found that these special lowest-twist coefficients can be written as functions of $\Delta \equiv \Delta_L$  
and $f_0 \sim \Delta_H/C_T$,  and thus all the model-dependent data can be fully absorbed into the central charge $C_T$. See \eqref{f0sum}.
In particular, the structure in lowest-twist limit is not altered by higher-curvature terms in the gravitational action beyond Einstein gravity.  
The higher-twist OPE coefficients, on the other hand, are generally contaminated by such higher-curvature terms.

In $d=2$,  the Virasoro algebra essentially determines all the related structures, but 
it is not clear a priori whether a similar algebraic approach can also work in higher dimensions. 
Instead of directly searching for a higher-dimensional generalization of the Virasoro algebra from the scratch, the 
holographic framework has provided us with a concrete starting point to gain some mileage and also develop initial intuition. 
Our holographic computations suggest that some version of a  higher-dimensional Virasoro symmetry may exist, at least 
in the lowest-twist and large central charge limits.   
While we shall not further discuss a general field-theoretic approach here, we believe 
revisiting some previous works \cite{Herzog:2013ed, CC1989, Huang:2013lhw}  
relating the stress tensor in higher-dimensional CFTs with the conformal anomaly central charges could be 
useful. 

On the other hand, we hope that the story on the gravity side is far from the end. 
Let us conclude by briefly mentioning some future problems.

{
\addtolength{\parskip}{1 ex}

$\bullet$ For generic $\Delta$, we have not been able to obtain analytic resummations of the 
lowest-twist $T^n$ contributions, and our series expansion typically has finite radius of convergence 
beyond which we have not been able to explore. It would be useful to try to numerically solve the PDEs and compare to the results obtained in this paper.
Moreover, numerical computations should help extracting lowest-twist data in odd dimensions, for which we have provided 
few analytic expressions.

$\bullet$ Because the stress-tensor contributions 
get mixed by the double-trace modes, which require an interior boundary condition to be fully determined, we 
do not know if the universality continues to hold when $\Delta$ is an integer.
While we do not expect double-traces can be universal, it would be 
interesting to see the non-universality explicitly, and to investigate universality of the poles (as a function of $\Delta$) in the double-trace contributions.

$\bullet$  We have assumed that two operators to be heavy to compute 
the two-point function in the black hole background.  A natural question one can ask is 
whether or not there is a universality away from the heavy limit.  
Implementing the method of geodesic Witten diagram \cite{Hijano:2015zsa} might shed light on this question.

$\bullet$ The results in this paper are valid only in the large central charge $C_T$ limit because we have ignored  loops.  
It would be interesting to study whether some form of universality for the lowest-twist coefficients remains after including $1/C_T$ corrections, 
and to know to what extent the universality starts to break down.  Unlike in $d=2$, the dimensions of $T^n$ 
operators are not generally protected and they should develop anomalous dimensions, which may indicate 
an obstacle to developing an algebraic approach at finite $C_T$.

$\bullet$ Although the case of the spherical black hole is more complicated and we have weaker results here, 
we still expect that the lowest-twist  coefficients are again universal (see appendix \ref{app:spherical}).  
In particular, we make a ``strong" conjecture in appendix \ref{app:spherical} which would be nice to prove, as it would 
imply that the high-temperature near-lightcone limit is universal in a larger region where one can separately move along the lightcone and away from it.

$\bullet$ It has been known that the gravitational shockwave geometry is insensitive 
to the higher-order curvature corrections \cite{Horowitz:1999gf}. 
We would like to understand better the relationship between a computation performed in 
a shockwave background and the results obtained in the black hole  background considered in this paper.  
We expect that there is a map between these two kinds of computations.

$\bullet$ It will be interesting to explore the similar lowest-twist universality 
and thermalization phenomenon in the context of higher-dimensional boundary/defect CFTs, either from field-theory or gravity side. 
A recent graphene-like conformal model has been found to allow explicitly 
marginal-dependent central charges \cite{Herzog:2017xha, Herzog:2018lqz} and thus might serve as a toy model to perform simpler perturbation.  

$\bullet$ For simplicity, we have ignored matter fields when solving 
the bulk field equation. It would be interesting to generalize 
the computation considered here to include matter fields and see
if  the lowest-twist universality persists in certain ways.  
To confirm our lowest-twist results using specific superconformal field 
theories such as $d=4$ $N=4$ Super-Yang-Mills or $d=3$ ABJM theory \cite{Aharony:2008ug}, the relevant matter fields 
should be included in the gravity action.   
\\
}

This last point, concerning the matter content of the bulk theory, deserves further comment.  We have made two important assumptions about the effective bulk Lagrangian we used. First, that any additional bulk fields can be integrated out to generate only local terms in the Lagrangian, and second that additional couplings between the probe field $\phi$ can be neglected.  

The former of these can be formalized as an assumption that there is a parameter in the CFT that allows one to make the extra bulk fields arbitrarily heavy.  To all orders in an expansion in this parameter, the effects of integrating out the extra bulk fields can be absorbed by local terms, so our prediction of universality should be thought of as a prediction for the terms in such a series expansion.  Optimistically, the long-distance near-boundary nature of the lowest-twist modes may mean that even bulk fields with large but finite masses can be effectively absorbed into local terms for the purposes of our calculation. It will be important to investigate such issues in more detail.

The second assumption is that additional bulk interactions of the probe field $\phi$ (such as $R \phi^2$ or $\phi^4$ term) can be neglected.  
This point can be addressed similarly to the first, by integrating out fluctuations of the bulk field $\phi$ around a background $\phi_0$, and considering the resulting effective action for $\phi_0$ at quadratic order.  If the mass of $\phi$ is large, integrating out its fluctuations generates local terms of the form
\be
\a \phi_0 \left(g_{\dots} \dots g_{\dots} R^{\mu_1 \nu_1 \rho_1\sigma_1} \dots R^{\mu_n \nu_n \rho_n \sigma_n} \nabla^{\alpha_1} \dots \nabla^{\alpha_{2s}}  \right) \phi_0 \ .
\ee
These terms generate additional contributions to the equation of motion for the bulk-to-boundary propagator $\Phi$.  However, we isolate the lowest-twist parts of $\Phi$ by taking $r\rightarrow$ with $u,w$ fixed (and $\Phi$ in the form in eq. (\ref{eq:LTlimitform})), and at infinite $r$, the Riemann tensor agrees with that of pure AdS.  Thus, the extra terms above reduce to
\be 
\label{higherdsphi}
\a \phi_0 (\nabla^2)^s \phi_0 \ .
\ee
With the addition of this new term, the equations of motion at large $r$ therefore read 
$\nabla^2\Phi= m^2 \Phi + \a (\nabla^2)^s \Phi$.  
Our previous solution $\Phi$ satisfies $\nabla^2 \Phi = \Delta_L(\Delta_L-d) \Phi$, so in particular its lowest-twist piece is still a solution, but with a ``renormalized''  mass:
\be
\Delta_L(\Delta_L-d) = m_{\rm eff}^2=  m^2 +  \a m^{2s} \ .  
\ee 
Since our conclusions about lowest-twist OPE coefficients were formulated in terms of the physical dimension $\Delta_L$, they should remain unchanged.  

In the above argument, higher-order terms $\sim \phi_0^n$ in the background $\phi_0$ can be neglected as they do not affect the equation of motion for the bulk-to-boundary propagator; by contrast, higher-order terms $\sim \phi^n$ in the original Lagrangian will affect the resulting effective Lagrangian for $\phi_0$ when the fluctuations of $\phi$ are integrated out.  
The form \eqref{higherdsphi} therefore captures the most general (local) higher-derivative interacting scalar theory, 
 assuming that one is able to expand in powers of $1/\Delta_L$. 
It would be interesting to know whether they hold more generally. 

\newpage

\subsection*{Acknowledgments}

We are grateful to Jared Kaplan for many useful discussions and in particular for emphasizing  
the utility of the simultaneous high-temperature near-lightcone limit.  
We also thank Ethan Dyer, Thomas Hartman, Kristan Jensen, Daliang Li, Jo\~ao Penedones, and Sasha Zhiboedov   
for sharing their insights. ALF and KWH were supported in part by the US Department of Energy Office of Science 
under Award Number DE-SC0015845  and in part by the Simons Collaboration 
Grant on the Non-Perturbative Bootstrap, and ALF in part by a Sloan Foundation fellowship.

\newpage
\appendix

\renewcommand{\theequation}{5.\arabic{equation}}
 \setcounter{equation}{0}

\section{Spherical Black Hole}
\label{app:spherical}
 
The structure becomes more complicated with a spherical black hole, but the general 
scheme is largely the same as what we have considered in the planar black hole case. 
To reduce repetition, in this appendix we focus on \\
(i) the field equation and corresponding change of variables;  \\
(ii) an explicit $d=4$ example;\\
(iii) a discussion of universal lowest-twist.   

\subsubsection*{{\it Preliminary Remarks:}}

A key point is that in the spherical black hole case, there are more $T^n$ operators that contribute to the 
conformal block decomposition than in the planar black hole limit.  The reason is that in the planar black hole limit, temperature 
is taken to infinity while the separation between the light operators is taken to zero, and many $T^n$ operators decouple in this limit.  
Consequently, it is not immediately clear how the universality of lowest-twist $T^n$ 
operators in the planar limit should generalize to spherical black holes. We will discuss a few different conjectures.  

The weakest conjecture is that the lowest-twist $T^n$ operator at each $J$ is universal, and we will provide compelling evidence for this.  
In our explicit computations in $d=4$ up to dimension 14, however, we see evidence for a stronger conjecture: the lowest-twist $T^n$ 
operators at each $n$ are universal.  At each $n\ge 2$, there are an infinite number of $T^n$ operators with the lowest 
possible twist $\tau_{\rm min}(n)=n(d-2)$, so this conjecture is much stronger than the weak version.  

Despite this fact, we find in our explicit computations that there are even more $T^n$ operators whose OPE coefficients depend only on $f_0$ and $\Delta_L$ than 
are accounted for by the strong conjecture, so perhaps an even stronger statement of universality holds.

We emphasize that the strong conjecture mentioned above would be useful for probing the heavy-light correlators in more detail.  
The reason is that in the near-lightcone, high-temperature limit where $\bar{z}$ is taken to be small with $\bar{z}^{\frac{d-2}{2}} f_0$ fixed, but $z$ is 
taken to be $\CO(1)$, the heavy-light correlator depends only on the $T^n$ operators fixed by the strong conjecture, and would depend on two free 
parameters $z$ and $\bar{z}$ independently.  By contrast,  in the planar black hole limit where both $z$ and $\bar{z}$ are taken to be small, the heavy-light 
correlator depends only on the combination $f_0 (z \bar{z})^{\frac{d-2}{2}} z^2$.

\subsection{Field Equation and Change of Variables}

We start with 
\be
\label{metricFHS}
ds^2= \big(1+r^2 f(r)\big) dt^2  + {dr^2 \over 1+ r^2 h(r) } + r^2 \sum_{i=1}^{d-1}d\Omega_i^2\ ,
\ee 
where $d\Omega_i^2$ with angular coordinates $(\theta_1, \theta_2\ , ... \theta_{d-1})$ is the metric on a unit $(d-1)$-sphere. 
In the following, we shall remove $(\theta_2\ , ... \theta_{d-1})$ dependence in the scalar field $\Phi$ with the 
help of the rotation symmetry and rename $\theta_1=\theta$. 
The scalar field equation can be written as  
\be 
\label{EoMsph}
&&\Big[{\big(1+a^2\big)\del^2_a+a\partial_a\over 1+r^2 f} + {\big(1-b^2\big)\partial^2_b
+\big({2\over b} - 3 b\big)\del_b\over r^2}+  \Big(1+r^2 h\Big) \del^2_r  \nn\\
&&~~+{ r^2 \Big( h \big( 8+ r^3 f'  \big)  +r\big( f'+h' \big)+f \big( 8+10 r^2 h +r^3 h' \big)
\Big)+6\over 2r (1 + r^2 f)} \del_r  \nn\\
&&~~+\big(d-4\big)\Big( {1+r^2 h\over r}\del_r+ {1 - b^2\over r^2 b}\del_b +\Delta \Big) \Big] \Phi (r, a, b)=\Delta (\Delta-4)\Phi (r, a, b) \ ,
\ee 
where 
\be
a=\sinh (t) \ , ~~~~~ b=\sin(\theta) \ .
\ee
The free solution, 
\be
\label{freesph}
\Phi_{\rm{free}}= \left({1\over 2}\frac{1}{ \sqrt{(1+r^2)(1+a^2) }-r\sqrt{1-b^2}}\right)^{\Delta} \ ,
\ee 
solves $\eqref{EoMsph}$ in the case of pure AdS.   Next, we define 
\be
(\hat a, \hat b)= r (a, b)  \ ,
\ee 
and consider the following change of variables:
\be
(r, a, b) \to (r, \hat a, \hat b) \to (r, \tilde{w}, \hat b) \ , 
\ee
where we introduce
\be
\label{wsph}
\tilde{w}^2= 1+ \hat a^2 + \hat b^2 \ .
\ee  
The relation betwen $\tilde{w}$ and $w$ defined in \eqref{definewsph} is 
\be
\lim_{r\to \infty~{\rm{with}}~\hat t, \hat \theta~{\rm{fixed}}} \tilde w=w \ .
\ee 
Namely, $w$ is the short-distance limit of $\tilde w$ in the boundary limit.  

 The variable $\tilde{w}$ in the present spherical black hole case is not directly suggested by the free 
solution \eqref{freesph}. Instead, it is largely suggested by the planar black hole 
case considered in the main text.  The reason for using $\tilde{w}$ is that the field equation becomes simpler 
in terms of variables $a$ and $b$. 
It is straightforward to rewrite the scalar field equation in terms of new 
variables $(r,\tilde{w},\b)$, but the full expression becomes bulky and we will not list it explicitly here.  

Considering a large $r$ expansion, one can write  
\be
\label{fh}
 f(r) = 1-{1\over r^d} \sum_{i=0,1,2,..} {f_i \over r^i} \ , ~~~ h(r)=  1- {1\over r^d} \sum_{i=0,1,2,..} {h_i \over r^i} \ .
\ee  
The standard AdS-Schwarzschild spherical black hole solution, $f(r)=h(r)=1- {f_0\over r^d}$, is recovered if removing higher-curvature corrections.
We expect the conformal block decomposition constrains the allowed powers in \eqref{fh}, but 
the analysis becomes more complicated in the spherical case as there are additional contributions from 
 inserting derivatives into any two stress tensors. 
We shall not go into such classification details here as the universality of the lowest-twist coefficients shall not 
depend on higher-order structures.

\subsection{Example}

Here we consider the spherical black hole generalization of section \ref{example}, where we took $d=4$ with  
\be
\label{fh4d}
 f(r) = 1- {1\over r^4} \sum_{i=0,4,8,...} {f_i \over r^i} \ , ~~~ h(r)=  1- {1\over r^4} \sum_{i=0,4,8,...} {h_i \over r^i} \ .
\ee   
There are stress-tensor and double-trace contributions: 
\be
 \Phi = \Phi_{\rm {free}}  G(r, \tilde{w} , \b)\ ,~~~ G(r, \tilde{w} , \b)= 1+G^{\rm{T}}(r, \tilde{w}, \b)+ G^{\rm{\phi}}(r, \tilde{w}, \b) \ . 
\ee 
The double-traces, $G^{\rm{\phi}}$, require an 
interior boundary condition to be fully determined, and we shall drop them in the following. 

After imposing the $\delta$-function boundary condition and the regularity at $\tilde{w}=1$,  
we find the stress-tensor solutions admit polynomial forms, similar to the planar black hole case:
\be
\label{GTsph}
&&G^{\rm{T}}(r, \tilde{w}, \b)={1\over r^4}\sum_{i=0,2,4,6,...}{G_i^{\rm{T}}(\tilde{w}, \b)\over r^i} \ , 
\ee 
with 
\be
G^T_{0}&=&\a^{(0)}_0(\tilde{w})+ \a^{(2)}_0(\tilde{w})\b^2\nn\\
&=& \sum_{i=-2}^{4} a_i \tilde{w}^i+\sum_{j=-2}^{2} b_j \tilde{w}^j  {\b}^2\ , \\
G^T_{2}&=&\a^{(0)}_2(\tilde{w})+ \a^{(2)}_2(\tilde{w})\y^2+ \a^{(4)}_2(\tilde{w})\b^4 \nn\\
&=& \sum_{i=-4}^{6} a_i \tilde{w}^i+\sum_{j=-4}^{4} b_j \tilde{w}^j  {\b}^2+\sum_{k=-4}^{2} c_k \tilde{w}^k  {\b}^4\ , \\   
G^T_{4}&=& \a^{(0)}_4(\tilde{w})+ \a^{(2)}_4(\tilde{w})\y^2+ \a^{(4)}_4(\tilde{w})\y^4+\a^{(6)}_4(\tilde{w})\y^6   \nn\\ 
&=&  \sum_{i=-6}^{8} a_i \tilde{w}^i+\sum_{j=-6}^{6} b_j \tilde{w}^j  {\b}^2+\sum_{j=-6}^{4} c_k \tilde{w}^k  {\b}^4+\sum_{l=-6}^2 d_l \tilde{w}^l  {\b}^6 \ ,
\ee 
and so on. One may observe a general pattern.
Note that  $1/r^6$, $1/r^{10}$, $1/r^{12}$, etc are allowed powers in the perturbative 
solutions \eqref{GTsph} in the spherical black hole case.  

The solution $G^T_{0}$ is the same as that in the planar black hole case, \eqref{GT0}-\eqref{GT0b}.  
We find the consistency with the conformal block decomposition again 
requires $h_0=f_0$ and higher-order $f$- and $h$-factors are not constrained. 
In the following, we shall set $h_0=f_0$, which also makes the expressions simpler.  

At the next order, we obtain 
\be
&&G^T_{2}= - {\Delta  f_0\over \tilde{w}^4} \Big(  \frac{ 9 \tilde{w}^{10}-15 \tilde{w}^8+12 \tilde{w}^6+244 \tilde{w}^4-544 \tilde{w}^2+224 }{560}\nn\\
&&~~~~~~~~~~~~~~~~~ -\frac{ 20 \tilde{w}^8-11 \tilde{w}^6-10 \tilde{w}^4+288 \tilde{w}^2-252 }{420}~\b^2 \nn\\
&&~~~~~~~~~~~~~~~~~~ +\frac{ 8 \tilde{w}^6+5 \tilde{w}^4-6 \tilde{w}^2+42  }{210}~\b^4 \Big) \ .
\ee 
This solution still corresponds to a single stress-tensor exchange. 
To go beyond, one must look at higher orders.

The full expression of $G^T_{4}$ however becomes cumbersome. Let us first list 
the leading and subleading large $\b$ part of $G^T_{4}$:\footnote{We find that only the zeroth-order in $\b$ part of solution $G^T_{4}$ contains factor $h_4$.}
\be
&&G^T_{4}=- \frac{\Delta  f_0 \left(4 \tilde{w}^8-2 \tilde{w}^4-3 \tilde{w}^2+21\right) }{105 \tilde{w}^6} \b^6 \nn\\ 
&&
~~~~~~~~~ +{\Delta f_0 \over \tilde w^6}  
\Big[\frac{ \left(7 \Delta ^2+6 \Delta +4\right) f_0+ 840 (\Delta -2)}{12600 (\Delta -2)}\tilde w^{10}  \nn\\ 
&&~~~~~~~~~~~~~~~~~~~~~ -\frac{160 (\Delta -1)- \left(7 \Delta ^2+6 \Delta +4\right) f_0}{2100 (\Delta -1)} \tilde w^8\nn\\ 
&&~~~~~~~~~~~~~~~~~~~~~ -\frac{5 \Delta - \left(7 \Delta ^2+6 \Delta +4\right) f_0}{600 \Delta} w^6  
+\frac{ 7 (9 \Delta +8) f_0-120}{3150}\tilde w^4 \nn\\ 
&&~~~~~~~~~~~~~~~~~~~~~ +\frac{ 7 (\Delta +1) f_0+320}{350} \tilde w^2  
-\frac{4}{5}\Big] \b^4+ G^T_{4} (\b^2, \b^0) \ .
\ee   
As we are mostly interested in the large $r$ limit (with variables $a, b$ fixed), next we simply list the 
effective contributions in the remaining $G^T_{4} (\b^2, \b^0)$ that survive in the boundary limit:\footnote{By sending $\Delta\to 0$, these stress-tensor contributions do not contribute 
in the boundary limit, but they remain non-zero at some finite $r$. The full scalar solution, including 
double-traces, should be trivial when $\Delta=0$.}
\be
&&\lim_{r\to \infty} G^T_{4} (\b^2, \b^0)\nn\\
&&=\Delta\Big(\frac{-5 f_0 }{112}  
+\frac{ {\cal A} f_0^2- {\cal B} f_4 }{25200 (\Delta -3) (\Delta -2)}\Big)    \tilde{w}^6 \b^2 \nn\\  
&&~~~ +\Delta\Big(\frac{149  f_0}{13440} 
+ {{\cal C} f_0^2- {\cal D} f_4- {\cal E} h_4 \over 201600 (\Delta -4) (\Delta -3) (\Delta -2)}   \Big)  \tilde{w}^8 +{\cal O} (r^6) \ ,  
\ee 
where 
\be
&&{\cal A}= 7 \Delta^2 (7-3 \Delta )+24\Delta+24 \ ,\\
&&{\cal B}= 40 (\Delta +1) (\Delta +2) \ , \\
&&{\cal C}= 7\Delta^2 (\Delta -3) (9 \Delta -32)   -88\Delta +144\ , \\
&&{\cal D}= 160 \Delta^2  (3-2 \Delta )+800\Delta  \ , \\
&&{\cal E}= 40 (\Delta +1) \big((\Delta -4) \Delta +24\big) \ .
\ee  
It is straightforward to go further and obtain higher-order solutions, albeit increasingly unwieldy.  

Let us finish this example by 
performing the conformal block decomposition and extracting explicit OPE coefficients. 
The coordinate transformations \eqref{eq:zzbarTOttheta} give   
\be
(a, b)= {1\over 2 \sqrt{y \bar y}} \Big(  y \bar y-1 ,  i (\bar y - y)\Big) \ , 
\ee  where $y=1-z$, $\bar y = 1-\bar z$. 
The coefficients $c_{\rm{OPE}}(4,2)$, $c_{\rm{OPE}}(8,0)$, $c_{\rm{OPE}}(8,2)$ and 
$c_{\rm{OPE}}(8,4)$ are the same as that obtained from a planar black hole.\footnote{In the spherical black hole 
case, one does not need to perform additional rescalings \eqref{rescaling}.}
After computing $G^T_{6}$, which will not be spelled out here, and performing the conformal block decomposition 
in the boundary limit, we find
\be
&&c_{\rm{OPE}}(10,0)=
{\Delta  (\Delta +1)\over 310464000 (\Delta -5) (\Delta -4) (\Delta -3) (\Delta -2)} \nn\\
&&~~~~~~~~~~~~~~~~~~~~~~ \times \Big[ 9  f_0^2  \Big(77 \Delta ^4-488 \Delta ^3+1207 \Delta ^2 -1028 \Delta +1840\Big) \nn\\
&&~~~~~~~~~~~~~~~~~~~~~~~~~~~~~~ + 160  \Big(137 f_4-65 h_4\Big) \Big(\Delta ^3-2 \Delta ^2+37 \Delta +90\Big)\Big] \ ,  
\ee
\be
&&c_{\rm{OPE}}(10,2)={\Delta\over 79833600 (\Delta -4) (\Delta -3) (\Delta -2)} \nn\\
&&~~~~~~~~~~~~~~~~~~~ \times\Big[ f_0^2\Big(187 \Delta ^4-552 \Delta ^3+901 \Delta ^2 +1012 \Delta +912\Big)\nn\\
&&~~~~~~~~~~~~~~~~~~~~~~~~~~~~~~~  + 800\Big(7 f_4+h_4\Big) \Big(\Delta ^3+6 \Delta ^2+11 \Delta +6\Big) \Big]\ ,  
\ee
\be
\label{subtwistsph}
&&c_{\rm{OPE}}(10,4)= { 9 \Delta ^3+8 \Delta ^2+31 \Delta +12 \over 2956800 (\Delta -3) (\Delta -2)} \Delta  f_0^2\ , 
\ee 
and the lowest-twist
\be
c_{\rm{OPE}}(10,6)= {33 \Delta^2-7 \Delta +4\over 38438400 (\Delta -2)} \Delta  f_0^2 \ . 
\label{eq:OPE106}
\ee  

As a reference, we list the lowest-twist coefficients at the next two levels:
\be
&&c_{\rm{OPE}}(12,8)=
\frac{286 \Delta^2 -157 \Delta+12}{8576568000 (\Delta -2)}\Delta f_0^2\ , 
\label{eq:OPE128}\\
&&c_{\rm{OPE}}(14,10)=\frac{325 \Delta ^2-229 \Delta +6}{219011240448 (\Delta -2)}\Delta  f_0^2 \ .
\label{eq:OPE1410}
\ee  

\subsubsection*{Remarks on subleading-twists}

Looking at the subleading-twist coefficient \eqref{subtwistsph}, one may wonder if in the spherical black hole case the 
subleading-twist coefficients are also universal.   
A direct higher-order computation provides some evidence:
\be
&&c_{\rm{OPE}}(12,6)={ \Delta  f_0^2 \over 72648576000 (\Delta -3) (\Delta -2)}\nn\\
&&~~~~~~~~~~~~~~~~~~\times \Big[7 \left(1001 \Delta ^4+3575 \Delta ^3+7310 \Delta ^2+7500 \Delta +3024\right) f_0\nn\\
&&~~~~~~~~~~~~~~~~~~~~~~~~~ +144 \left(44 \Delta ^3-37 \Delta ^2+71 \Delta +12\right)\Big] \ , 
\label{eq:OPE126} \\
&&c_{\rm{OPE}}(14,8)=  
\frac{\Delta  f_0^2}{31287320064000 (\Delta -3) (\Delta -2)} \nn\\
&&~~~~~~~~~~~~~~~~~~~\Big[ 51 \left(3003 \Delta ^4+6032 \Delta ^3+9029 \Delta ^2+7148 \Delta +2688\right) f_0\nn\\
&&~~~~~~~~~~~~~~~~~~~~~~~~ + 200 \left(559 \Delta ^3-746 \Delta ^2+709 \Delta +54\right)\Big] \ .
\label{eq:OPE148} 
\ee
On the other hand, we find the corresponding sub-sub-leading twist coefficients both have the following $f$-dependence: 
\be
\label{eq:MoreOPEs0}
&& c_{\rm{OPE}}(12,4)= c_{\rm{OPE}}(12,4) (f_0^2, f_0^3, f_0 f_4)\ , \\
&& c_{\rm{OPE}}(14,6)=c_{\rm{OPE}}(14,6) (f_0^2, f_0^3, f_0 f_4) \ ,
\label{eq:MoreOPEs}
\ee with no linear in $f_4$ term. The factor $h_4$ starts to appear in higher twists coefficients $c_{\rm{OPE}}(12,2)$, $c_{\rm{OPE}}(14,4)$.  
While exploring these subleading corrections is not the main focus of the present work, we will comment on them more with some conjectures later. 

Different from the planar black hole case where the sub-leading large $\y$ solutions 
generally are not universal, in the  present spherical black hole case we find 
factors $f_i/h_i$ with $i>0$ enter the scalar solution only starting at sub-sub-leading large $\b$ order.  
Moreover, solutions at leading $\b$ depend on the $f_0$ 
only through $f_0^1$ while the sub-leading $\b$ solutions depend on $f_0$ 
through both $f_0^1$ and $f_0^{\#}$, with certain powers $\#$ that increase with the order of $1/r$.  
The interpretation is that the single stress-tensor contribution continues to affect higher-order scalar solutions in the spherical black hole case. 
On the other hand, there should not have $\sim f_0^1$ contribution in the multi-stress-tensor conformal blocks.
This suggests that probing the lowest-twist in the spherical black hole case requires a large $\b$ limit keeping the $sub$-$leading$ contribution.

\subsection{Universal Lowest-Twist}
 
We focus on $d=4$ for concreteness. It should be straightforward to generalize to other dimensions. 
Similar to a large $\y$ in the planar black hole case, here we consider 
\be
\label{largeb}
\b \to \infty \ .
\ee
Defining 
\be 
u={\b \over r} \ , 
\ee which gives simply $u=b$, 
and performing change of variables from $(r,\tilde{w},\b)$ to $(r,\tilde{w},u)$, 
the limit \eqref{largeb} can be achieved by taking a large $r$ with $\tilde{w}$, $u$ fixed.  
As mentioned earlier, we expect that the 
information of the lowest-twist is contained in the $sub$-$leading$ large $\b$ limit of the scalar 
perturbative solutions. (We find that the leading $\b$ limit picks up contributions $\sim f_0^1$, corresponding to a single stress-tensor.)

To include the sub-leading contributions, we write\footnote{Perturbative structures read
\be
&& P=P_2(\tilde{w}) u^2+ P_4(\tilde{w}) u^4+ P_6(\tilde{w}) u^6 +P_8(\tilde{w}) u^8 ...  \ , \\
&& Q=Q_2(\tilde{w})+ Q_4(\tilde{w}) u^2+ Q_6(\tilde{w}) u^4 + Q_6(\tilde{w}) u^6...  \ .
\ee   
}
\be
\label{sphGQ}
&&\lim_{r\to \infty} G^{T} (r,\tilde{w}, u) = {P(\tilde{w},u)\over r^2}+ {Q(\tilde{w},u)\over r^4} +{\cal O}({1\over r^6})  \ .
\ee
Starting with \eqref{EoMsph} and performing the change of variables, we 
find that the leading- and sub-leading reduced field equations can be symbolically written as\footnote{Since we 
require the sub-leading term, we need to determine the power of ${1\over r}$ in $f(r)$ and $h(r)$ beyond the leading order, $1/r^4$.  
We do not find a non-trivial solution consistent with the conformal block decomposition when $f(r)$ or $h(r)$ has $1/r^5$, $1/r^6$, or $1/r^7$ structure, and 
thus we take the next order to be $f_4/r^8$, $h_4/r^8$ in $f(r)$, $h(r)$, respectively. 
}
\be
\label{spherical proof0}
F_1\Big(f_0, \Delta,  P(\tilde{w},u) \Big) &=& 0 \ , \\
\label{spherical proof}
F_2\Big(f_0, h_0, \Delta, P(\tilde{w},u), Q(\tilde{w},u) \Big) &=& 0 \ ,
\ee respectively. 
The consistency with the conformal block decomposition requires $h_0=f_0$.
The explicit forms of $F_1$ and, in particular, $F_2$ are unwieldy and  thus will  not be spelled out here.   
The main point is that the above two reduced equations are both  protected: 
they do not depend on $f_i$ or $h_i$ with $i>0$.
We find the sub-sub-leading reduced equation instead  depends on higher-order $f_i/h_i$.

Next we shall discuss how \eqref{spherical proof0}, \eqref{spherical proof} suggest the universal lowest-twist with a spherical black hole.
One first solves for $P$ using the reduced equation $F_1$, \eqref{spherical proof0}. 
The $\delta$-function boundary condition together with  
the regularity at $\tilde{w}=1$ fix two integration constants and thus $P$ depends on $f_0$ only. 
Plugging $P$ into $F_2$, \eqref{spherical proof}, one next solves for $Q$. 
The same boundary conditions again fix two integration constant, and we conclude $Q$ is also protected.  

The more complicated part is to see why a protected $Q$ implies universal lowest-twist. 
Let us first discuss what it means by having a protected $P$.
We find that $P$ depends on $f_0$ linearly.\footnote{This may be verified using explicit solutions or 
by deriving a general recursion relation for $P$.} 
The solution $P$ therefore belongs to the single stress-tensor contribution. 
As mentioned, in the spherical black hole case, the single stress-tensor contribution 
propagates to higher-order scalar solutions but in the multi-stress-tensor conformal blocks all 
linear-in-$f_0$ terms must cancel out. To probe the lowest-twist we shall look at the subleading $\b$.  

Denote the boundary limit of solutions $P_i$ and $Q_i$ as
\be
&& \bar P_i (\tilde{w})= \lim_{\tilde{w}\to \infty}   P_i (\tilde{w})  \ , \\
&& \bar Q_i (\tilde{w})=\lim_{\tilde{w}\to \infty}   Q_i (\tilde{w})  \ . 
\ee
Say we already know that $c_{\rm{OPE}}(4,2)$ and $c_{\rm{OPE}}(8,4)$ are protected and  
want to see if the next level's lowest-twist coefficient, $c_{\rm{OPE}}(10,6)$, is protected without 
explicitly computing the scalar solution. First we may look at the leading large $\tilde{w}$ of the subleading large $\b$ limit of  $G_6$: $\bar Q_6(\tilde{w})u^6  \sim \tilde{w}^4 u^6$.
The structure $\tilde{w}^4 \b^6/r^{10}$ is mapped to $z^2 \bar z^2 (z - \bar z)^6\sim z^8 \bar z^2$ in the boundary limit. 
The coefficient $\bar Q_6$ however is not the only coefficient associated  with $z^2 \bar z^2 (z - \bar z)^6$.  In fact,  
all leading- and subleading-$\b$ coefficients (in total eight coefficients at this level) are linked to $z^2 \bar z^2 (z - \bar z)^6$.
Since $\bar P_i, \bar Q_i$ are protected, as indicated by \eqref{spherical proof0} and \eqref{spherical proof}, the 
 coefficient of $z^2 \bar z^2 (z - \bar z)^6$ in the boundary-to-boundary correlator is protected. 
On the other hand, the coefficient of $z^2 \bar z^2 (z - \bar z)^6$ from the conformal block decomposition is 
\be
\label{mix}
\Big(\frac{56}{33}c_{\rm{OPE}}(4,2)+\frac{147}{26}c_{\rm{OPE}}(8,4)+c_{\rm{OPE}}(10,6) \Big) z^2 \bar z^2 (z - \bar z)^6 \ .
\ee 
We see $c_{\rm{OPE}}(10,6)$ is protected.  

Let us go on and ask if the next level's lowest-twist coefficient, 
$c_{\rm{OPE}}(12,8)$, is universal. At this level, we may focus on the structure 
$ z^2 \bar z^2 (z-\bar z)^8\sim z^{10} \bar z^2$, which is linked to a combination of the protected coefficients 
$\bar P_i, \bar Q_i$ with $i=2,4,6,8,10$. On the other hand, the coefficient of $ z^2 \bar z^2 (z-\bar z)^8$ from the 
conformal block decomposition is
\be
\Big(\frac{225}{143} c_{\rm{OPE}}(4,2)+\frac{756}{65} c_{\rm{OPE}}(8,4)+\frac{162}{17} c_{\rm{OPE}}(10,6)+c_{\rm{OPE}}(12,8) \Big)z^2 \bar z^2 (z-\bar z)^8 \ .
\ee 
We see $c_{\rm{OPE}}(12,8)$ is also protected.  

One should be able to identify an all-order pattern and give a general derivation but we have not exhaustively 
explored the detailed structures in the spherical black hole case.   We hope the above 
analysis, including the change of variables, could be useful.  Note that 
the mixed coefficients described above make it a more complicated task to directly extract a specific lowest-twist coefficient.  
It would be useful to develop an algorithm to effectively compute lowest-twist coefficients with a spherical black hole.

\subsection{Conjectures}

The results in the previous subsections are consistent with a few possible 
different conjectures for which $T^n$ operators are universal in the {\it spherical black hole} case.  
Here we will mainly discuss two such conjectures, a weak one and a strong one, though we will note 
that the data suggest that perhaps even the strong version is not as strong as it could be.
\\

$\bullet$ {\it The weak conjecture: 
For each spin $J$, the lowest-twist multi-stress-tensor  
operator for that spin is universal. Namely,    
\be
c_{\rm{OPE}}\big(\tau_{\rm min} =2(d-2),J\big)= f_0^2 {\cal H}_{J}(\Delta_L) ~~~J>2  \ . 
\ee 
The function ${\cal H}_{J}$ is independent of higher-curvature parameters. } 
\\

 For $J=2$, the only $T^n$  operator is $T^{\mu\nu}$ itself.  For $J>2$, the lowest-twist operators have 
two stress tensors and $J-4$ derivatives, suitably anti-symmetrized to make a primary operator.  
There is one such operator at every even $J\ge 4$, and its twist is just $2(d-2)$.  We show the 
schematic form of the possible $T^n$ operators one can make at each twist $\tau$ and spin $J$ in Fig. \ref{fig:ConjectureTable}.

In the previous subsection, we have discussed some potential strategies toward finding a general 
proof of the weak conjecture, which is supported also by some explicit computations \eqref{eq:OPE106}, \eqref{eq:OPE128}, \eqref{eq:OPE1410}. 
\\

$\bullet$ {\it The strong conjecture:  
For each number $n$ of stress tensors, all the multi-stress-tensor  
operators with the lowest allowed twist for that $n$ are universal.\footnote{The lowest allowed twist is $\tau_{\rm min}(n) = n(d-2)$. 
The strong conjecture is equivalent to the weak conjecture at $n=2$; the universality at $n=1$ is  trivial.   
} }  
\\

The equation \eqref{f0sum}, for which we gave a general derivation in the planar limit, then corresponds 
to the spherical black hole case's strong conjecture.  Note that there would be no difference between the 
strong and weak versions in the planar limit where only one $T^n$ can contribute at each $n$ and thus label $J$ is redundant.  

This strong form includes {\it all} the operators in Fig. \ref{fig:ConjectureTable}.  Any primary operators made 
from partial derivatives $\partial^\mu$ and stress tensors $T^{\mu\nu}$ without contracting any indices satisfy this stronger lowest-twist condition.

It would be interesting if the strong conjecture in the spherical black hole case were true, because it  
constrains  all the operators that have contributions  in the generic high-temperature, near-lightcone limit with 
small $\bar z$ but fixed $f_0 \bar{z}^{\frac{d-2}{2}}$ and fixed $z \sim \CO(1)$, as one can see from the small $\bar{z}$ form of the conformal blocks in (\ref{eq:LCBlock}).\footnote{We are assuming that 
 the coupling of gravitons to the heavy states is still parameterized by $G_N \Delta_H$, so that a diagram with $n$ gravitons is proportional to $\sim (G_N \Delta_H)^n \sim f_0^n$.
}
 By contrast, in the planar black hole limit, both $z$ and $\bar{z}$ are small, so hypergeometric 
function in (\ref{eq:LCBlock}) reduces to 1 and the heavy-light correlator depends on $z$ and $\bar{z}$  through the combination 
\be
\left[  (z \bar{z})^{\frac{d-2}{2}} \bar{z}^2\right]^n
\ee rather than through $z$ and $\bar{z}$ independently.

\tikzset{ 
table/.style={
  matrix of nodes,
  row sep=-\pgflinewidth,
  column sep=-\pgflinewidth,
  nodes={rectangle,draw=black,text width=4em,align=center},
  text depth=1.25ex,
  text height=2.5ex,
  nodes in empty cells
},
row 1/.style={nodes={fill=gray!10}},
column 1/.style={nodes={fill=gray!10}}
}

\begin{figure}[t!]
\begin{center}
\begin{tikzpicture}

\matrix (mat) [table]
{
& 2  & 4 & 6 & 8 & 10  \\
$(d-2)$ & $T$ & - & - & - & -  \\
$2(d-2)$ & -  & $T^2$ & $\partial^2 T^2$ & $\partial^4 T^2$ & $\partial^6 T^2$ \\
$3(d-2)$ & -  & - & $ T^3$ & $\partial^2 T^3$ & $\partial^4 T^3$ \\
$4(d-2)$ & -  & - & - & $ T^4$ &$\partial^2 T^4$ \\
};
\draw (mat-1-1.north west) -- (mat-1-1.south east);
\node at ([xshift=-7pt,yshift=-4.5pt]mat-1-1) {$\tau$};
\node at ([xshift=7pt,yshift=4.5pt]mat-1-1) {$J$};
\end{tikzpicture}
\end{center}
\caption{Table of schematic form of $T^n$ operators at fixed twist $\tau$ and spin $J$.  The derivatives $\partial$ should 
be applied to the products of $T$ in combinations that form primary operators.  No indices are contracted, as that would decrease the spin and increase the twist. 
}
\label{fig:ConjectureTable}
\end{figure}

The evidence for the strong conjecture is the explicit form of the OPE coefficients 
calculated in the previous subsections in $d=4$.  At $\Delta_T=4$ there is only the stress tensor 
itself and at $\Delta_T=8$, there is no room for any partial derivatives in the operator so both the strong and weak form agree.  
At $\Delta_T=10$, the only new primary we can make without contracting indices is schematically $\sim \partial^2 T^2$, which is a lowest-twist operator 
according to both the weak and strong conjectures.  Its OPE coefficient appears in \eqref{eq:OPE106} and indeed depends only on $f_0$ and $\Delta$.  
At $\Delta_T=12$, the primary operators we can make without contracting indices are schematically $\sim T^3$ and $\sim \partial^4 T^2$.  
Only the strong conjecture applies to $T^3$; the weak form does not apply since $\sim T^3$ has twist 6 and spin 6 whereas $\sim \partial^2 T^2$ 
has twist 4 and spin 6, i.e. the same spin but lower twist.  Looking at (\ref{eq:OPE128}) and (\ref{eq:OPE126}) 
for $\sim \partial^4 T^2$ and $\sim T^3$ respectively, we see that again both depend only on $f_0$ and $\Delta$.

A potentially important point, though, is to observe  
that $c_{\rm OPE}(12,6)$ {\it also} gets a contribution from an operator $\sim \partial^4 T^2$ with two contracted indices.  
This operator can be distinguished in (\ref{eq:OPE126}) from $\sim T^3$ since the former is proportional to $f_0^2$ whereas the latter 
is proportional to $f_0^3$.  Interestingly, neither the weak nor the strong conjecture applies to this operator, yet its OPE coefficient depends only on $f_0$ and $\Delta$.

At $\Delta_T=14$, the possible lowest-twist operators to consider are $\sim \partial^2 T^3$ and $\partial^6 T^2$, with 
spin $J=8$ and $J=10$, respectively.  From (\ref{eq:OPE1410}) and (\ref{eq:OPE148}), we see that both of their OPE 
coefficients depend only on $f_0$ and $\Delta$. Lower $J$ operators, however, depend on $f_4$, as indicated in \eqref{eq:MoreOPEs0} and \eqref{eq:MoreOPEs}.  

It would be interesting to study more OPE coefficients in order to test the strong conjecture more thoroughly, and to 
potentially uncover if an even stronger statement of universality holds.  We leave such investigations to future work.

\section{Some Explicit Solutions}
\label{app:sol}

As a reference, we here list the explicit perturbative solutions at 
level ${\cal O}({1\over r^8})$ in the planar black hole case with background \eqref{bg1}.

\be
G^T_{4}:~
\label{a-4}
&&a_{-4}=\frac{\Delta}{50}   (\Delta +1) \left(f_0+h_0\right){}^2 \ , 
\ee
\be
&&a_{-2}= -\frac{ \Delta }{900} \Big[3 (6 \Delta +37) f_0^2+4 (9 \Delta +17) h_0 f_0\nn\\
&&~~~~~~~~~~~~~~~~~~ +18 \Delta  h_0^2+57 h_0^2
+100 f_4+100 h_4\Big]\ ,  
\ee
\be
&&a_0= {1\over 1440} \Big[\left(-12 \Delta ^2+108 \Delta +57\right) f_0^2+2 (20 \Delta +3) h_0 f_0+12 \Delta ^2 h_0^2\nn\\
&&~~~~~~~~~~~~~~~~+12 \Delta  h_0^2-63 h_0^2
+40 (2 \Delta +3) f_4+80 \Delta  h_4-120 h_4\Big]\ ,  
\ee
\be
&&a_2= {1\over 5040 (\Delta -1)}\Big[- 20 h_4 \Delta^2 
-15 h_0^2 \Delta ^2+60 h_0^2 \Delta 
+ 80 h_4 \Delta\nn\\ 
&&~~~~~~~~~~~~~~~~~~~~~~~~~ +3 \left(46 \Delta ^2-51 \Delta -12\right) f_0^2 -36 h_0^2\nn\\
&& ~~~~~~~~~~~~~~~~~~~~~~~~~  + 20 \left(8 \Delta ^2-11 \Delta +18\right) f_4  
- 480 h_4\nn\\ 
&&~~~~~~~~~~~~~~~~~~~~~~~~~ +\left(17 \Delta ^2-47 \Delta +72\right) f_0 h_0\Big]\ ,
\ee
\be
&&a_4= {1\over 10080 (\Delta -2) (\Delta -1)}\nn\\
&&~~~~~~~~~\times \Big[-7 h_0^2 \Delta ^4+41 h_0^2 \Delta ^3
- 20 h_4 \Delta ^3\nn\\  
&&~~~~~~~~~~~~~~~~ -67 h_0^2 \Delta ^2
+ 60 h_4 \Delta^2 
+24 h_0^2 \Delta 
- 400 h_4 \Delta \nn\\ 
&&~~~~~~~~~~~~~~~~ +\left(56 \Delta ^4-184 \Delta ^3+239 \Delta^2-120 \Delta -72\right) f_0^2\nn\\
&&~~~~~~~~~~~~~~~~ + 40 \left(4 \Delta ^3-3 \Delta ^2-\Delta +6\right) f_4 
-480 h_4 
-36 h_0^2\nn\\
&&~~~~~~~~~~~~~~~~ +\left(14 \Delta ^4-95 \Delta ^3+101 \Delta ^2-2 \Delta +120\right) f_0 h_0\Big]\ , 
\ee
\be
&&a_6={\Delta \over 50400 (\Delta -3) (\Delta -2) (\Delta -1)}\nn\\
&&~~~~~~~~~\times \Big[ 7 h_0^2 \Delta ^4-65 h_0^2 \Delta^3
-40 h_4 \Delta^3\nn\\ 
&&~~~~~~~~~~~~~~ +160 h_0^2 \Delta^2 
+120 h_4 \Delta^2 
-120 h_0^2 \Delta 
-800 h_4 \Delta\nn\\ 
&&~~~~~~~~~~~~~~ +2 \left(56 \Delta ^4-310 \Delta ^3+605 \Delta ^2-390 \Delta -96\right) f_0^2 \nn\\
&&~~~~~~~~~~~~~~ -72 h_0^2
+40 \left(8 \Delta ^3-9 \Delta ^2-11 \Delta +6\right) f_4 
-960 h_4\nn\\ 
&&~~~~~~~~~~~~~~ +\left(-56 \Delta ^4+335 \Delta ^3-845 \Delta ^2+740 \Delta +336\right) f_0 h_0 \Big]\ , 
\ee
\be
\label{a8}
&&a_8={\Delta \over 201600 (\Delta -4) (\Delta -3) (\Delta -2)} \nn\\
&&~~~~~~~~~ \times
\Big[4 \left(28 \Delta ^4-176 \Delta ^3+349 \Delta ^2-216 \Delta -54\right) f_0^2\nn\\
&&~~~~~~~~~~~~~~~ - 40 \big(\Delta +1\big) \Big(\big(20\Delta -8 \Delta^2 \big) f_4+\big(\Delta ^2-4 \Delta +24\big) h_4\Big)\nn\\ 
&&~~~~~~~~~~~~~~~ -4 \big(14 \Delta^4-89 \Delta^3+221 \Delta^2-224 \Delta -108\big) f_0 h_0\nn\\
&&~~~~~~~~~~~~~~~ +\big(7 \Delta ^4-65 \Delta ^3+160 \Delta ^2-120 \Delta -72\big) h_0^2
\Big]\ , 
\ee 
and 
\be
\label{b-4}
&&b_{-4}= \frac{\Delta}{25} (\Delta +1) f_0 \left(f_0+h_0\right)\ , 
\ee
\be
&&b_{-2}=-\frac{\Delta}{180}  \left(19 f_0^2+h_0 f_0
+ 20 f_4\right)\ ,  
\ee
\be
&&b_0=- \frac{1}{360} \Big[\left(6 \Delta ^2+13 \Delta +16\right) f_0^2+(\Delta -16) h_0 f_0
+20 (\Delta +2) f_4\Big]\ ,  
\ee
\be
&&b_2= {1\over 840 (\Delta -1)}\Big[-\left(14 \Delta ^3-10 \Delta ^2+19 \Delta +12\right) f_0^2\nn\\
&&~~~~~~~~~~~~~~~~~~~~~~~+\left(4 \Delta ^2+19 \Delta +16\right) h_0 f_0
- 20 \left(\Delta ^2+3 \Delta +2\right)f_4\Big] \ ,  
\ee
\be
&&b_4=-{\Delta\over 5040 (\Delta -2) (\Delta -1)} \Big[4 \left(7 \Delta ^3-12 \Delta ^2+10 \Delta +4\right) f_0^2\nn\\
&&~~~~~~~~~~~~~~~~~~~~ -\left(7 \Delta ^3-13 \Delta ^2+52 \Delta +32\right) h_0 f_0 
+ 40 \left(\Delta ^2+3 \Delta +2\right) f_4\Big]\ ,  
\ee
\be
\label{b6}
&&b_6= {\Delta \over 25200 (\Delta -3) (\Delta -2)} \Big[-2 \left(14 \Delta ^3-31 \Delta ^2+14 \Delta +4\right) f_0^2\nn\\
&&~~~~~~~~~~~~~~~~~~~~ +\left(7 \Delta ^3-13 \Delta ^2+52 \Delta +32\right) h_0 f_0
- 40 \left(\Delta ^2+3 \Delta +2\right) f_4\Big]\ ,  
\ee 
and the universal part
\be
\label{c-4}
&& c_{-4}= \frac{f_0^2}{50} \Delta  \big(\Delta +1\big) \ , \\
&& c_{-2}= \frac{f_0^2}{450} \Delta  \big(9 \Delta +8\big)\ , \\
&& c_0=\frac{f_0^2}{600} \big(\Delta  (7 \Delta +6)+4\big) \ ,\\
&& c_2= \frac{\Delta  \big(\Delta  (7 \Delta +6)+4\big) f_0^2}{2100 (\Delta -1)}\ , \\
\label{c4}
&& c_4=\frac{\Delta  \big(\Delta  (7 \Delta +6)+4\big) f_0^2}{12600 (\Delta -2)} \ ,
\ee  
where $c_4$ determines the lowest-twist OPE coefficient.  
The conformal block decomposition requires $h_0= f_0$ but we have kept $h_0$ explicitly above.  

It is straightforward to obtain solutions to much higher orders, given the computation scheme discussed in this paper.  
But we emphasize that showing the universality of the lowest-twist coefficients does not require knowing these higher-order solutions.  

\newpage

\bibliographystyle{utphys}
\bibliography{LowestTwistBib}

\end{document}